# Ideometrics: The Science of Generating, Evaluating and Prioritising Ideas

Igor Rudan[1,2,3] and Aziz Sheikh[2]

[1]*Centre for Global Health, Usher Institute, The University of Edinburgh, UK*
[2]*Nuffield Department of Primary Care Health Sciences, University of Oxford, UK*
[3]*Green Templeton College, University of Oxford, UK*

## Abstract

In an age of exponential information growth, policy makers could benefit from an approach that would consistently generate, evaluate, and prioritise new ideas to help address the key challenges facing humanity. Across science, policy, business, and society, decision-makers are overwhelmed with competing concepts, hypotheses, proposals and agendas. Yet the approaches and tools we use to prioritise these ideas remain fragmented across disciplines and inconsistently applied. This paper introduces and outlines a new integrative field — Ideometrics. It is the systematic science of generating, evaluating, and prioritising ideas, based on the brain's proposed "sense of ideas" and "assigning value to information". Drawing on the structure based on those three fundamental aims, our work maps out the landscape of more than 70 methodologies that were applied to address those three aims across time, disciplines, and epistemic paradigms. The part on generating ideas synthesises classical creativity techniques, participatory approaches and computational innovations; the part on evaluating ideas explores the wide-ranging methods developed to this end; and the part on prioritising ideas assesses different prioritisation techniques. A central argument of this paper is that although these methods arose separately and independently, most often isolated by disciplinary boundaries, they share remarkable conceptual and structural similarities. Recurring features include creative ideation, aggregation of input, balancing of subjective and objective criteria, embedded value assumptions and iterative refinement. There is also a more recent trend of application of statistical inference and introduction of artificial intelligence, making Ideometrics increasingly quantitative, testable, replicable and digital. These insights suggest the feasibility and necessity of unifying these traditions under a common scientific framework. To this end, we propose several next steps for establishing Ideometrics as a rigorous and practically useful field of scientific enquiry. These include: (1) a bibliometric analysis to measure the scientific footprint of each of the >70 methodologies; (2) describing and evaluating the scientific approaches to be used; (3) the development of formal reporting guidelines to standardise Ideometrics studies and allow for future meta-analyses; (4) empirical comparisons between methods applied to the same problem sets; and (5) the creation of a user-friendly software platform that enables transparent, democratic, and replicable ideation workflows to support evidence-informed prioritisation of ideas across fields. We believe that this is the first systematic attempt to document, compare, and synthesise the full spectrum of methodologies used to generate, evaluate and prioritise ideas. While the task can hardly ever be completed, the framework proposed here is flexible and designed to evolve. Ideometrics aims to assist individuals, institutions, and societies to focus their limited capacity, time and resources on the ideas that could have the most positive future impact in a way that is transparent, inclusive, and analytically sound.

**Introduction: The Birth of Ideometrics – A Unified Science of Ideas**

Throughout human history, the capacity to generate, evaluate, and prioritise ideas has been at the heart of individual creativity, collective progress, and civilisational development [**1**]. Whether the context was philosophical debate in ancient Athens, medical diagnosis in Renaissance Italy, engineering problem-solving during the Industrial Revolution, or policymaking in modern democracies, the underlying process remained the same: faced with multiple competing ideas or possibilities, humans have always needed to decide which ideas are most worthy of pursuit [**2**]. In more traditional cultures, the emphasis was on ancestral traditions and folklore, the experiences of elders, intuition, revelation and spiritual enlightenment [**3,4**].

What differs across epochs, however, is the sophistication of the methods and processes used for such decisions. Across millennia, very many informal and formal techniques have been devised to aid idea generation, and we try to compile them in **Table 1**. They span across all human activities in their attempts to offer structured and replicable approaches to generating, evaluating and prioritising ideas, aiming to assist societies to allocate scarce time, funding, and attention to the most promising plans.

Yet despite their ubiquity and centrality across all domains, from science, medicine, and philosophy, to business, governance, and the arts, these methods and processes have rarely been treated as components of a unified scientific field. **Table 1** clearly shows that they have instead been dispersed across disciplines, each evolving its own tools, terminology, and criteria. Historically, their evolution was driven by local polymaths, rather than reductionist approaches, and their uptake and persistence likely depended on a degree of serendipity in surrounding contexts and circumstances. In isolation from each other, economists developed cost-effectiveness analysis, psychologists developed creativity techniques, engineers built optimisation models, political scientists refined deliberative forums, information theorists examined knowledge flows, and the funders in global health used the Child Health and Nutrition Research Initiative (CHNRI) method. But all of them were, essentially, trying to achieve the same goal in their own area of interest, using the same basic approach, that led them from developing, to evaluating, and then prioritising ideas.

As a result, many of the underlying intellectual challenges, such as how to quantify the potential of an idea, balance expert versus public input, or prioritise in the face of uncertainty, have been approached in parallel, but rarely synthesised. This fragmentation is precisely what motivates the emergence of a new integrative discipline that we have named Ideometrics: the science of generating, evaluating, and prioritising ideas. Ideometrics is not intended as a reductive or technocratic pursuit, but rather as a transdisciplinary framework that recognises the diversity of human knowledge systems while seeking coherence in how we work with ideas across them. Just as bibliometrics unified the study of scientific publications [**5**], and econometrics structured quantitative reasoning in economics [**6**], Ideometrics aspires to provide a systematic lens through which the full life cycle of ideas - conception, scrutiny, and their elevation and spread - can be understood and improved.

**Table 1.** The landscape of approaches, methods and tools that have been historically used to generate, evaluate and prioritise ideas.

| | | |
|---|---|---|
| 1. IDEA GENERATION METHODS<br><br>(approaches used to produce novel ideas, concepts or hypotheses) | 1.1. Individual and Cognitive Techniques | 1.1.1. Stream of consciousness and freewriting (William James, 1890 [**14**]; Peter Elbow, 1973 [**17**])<br>1.1.2. Theory of Inventive Problem Solving – TRIZ (Genrich Altshuller, 1946 [**18**])<br>1.1.3. Morphological analysis (Fritz Zwicky, 1957 [**20**])<br>1.1.4. Lateral thinking (Edward de Bono, 1967 [**22**])<br>1.1.5. SCAMPER technique (Robert F. Eberle, 1971 [**23**])<br>1.1.6. Mind mapping (Tony Buzan, 1974, 1993 [**24,25**])<br>1.1.7. Heuristics and biases framework (Amos Tversky and Daniel Kahneman, 1974 [**26**])<br>1.1.8. Six Thinking Hats (Edward de Bono, 1985 [**28**]) |
| | 1.2. Group-Based and Social Approaches | 1.2.1. Brainstorming (Alex F. Osborn, 1942, 1953 [**29,30**])<br>1.2.2. Focus groups and in-depth interviews (Robert K. Merton, Marjorie Fiske, Patricia L. Kendall, 1956 [**31**])<br>1.2.3. Delphi technique (Norman Dalkey and Olaf Helmer, RAND Corporation, 1963 [**34**])<br>1.2.4. Nominal Group Technique (Andre L. Delbecq and Andrew H. Van de Ven, 1971 [**35**])<br>1.2.5. World Café (Juanita Brown and David Isaacs, 1995 [**37**])<br>1.2.6. Open Space Technology (Harrison Owen, 1997 [**38**])<br>1.2.7. InnoCentive (Alpheus Bingham, Aaron Schacht, and Dwayne Spradlin, 1998, 2011 [**39**])<br>1.2.8. James Lind Alliance (Nick Partridge and John Scadding, 2004 [**40**], with Iain Chalmers)<br>1.2.9. Child Health and Nutrition Research Initiative – the CHNRI method (Igor Rudan, 2006, 2008 [**41,42**])<br>1.2.10. IdeaScale (Vivek Bhaskaran and Rob Hoehn, 2009 [**44**]) |
| | 1.3. Design and Innovation Frameworks | 1.3.1. Human-Centered Design (John E. Arnold, 1958 [**45**]; Donald A. Norman, 1988 [**46**]; IDEO.org, 2009 [**47**])<br>1.3.2. Design Thinking (Herbert Simon, 1969 [**48**]; Tim Brown, 2008 [**50**])<br>1.3.3. Agile ideation sprints (Ken Schwaber and Jeff Sutherland, 1995 [**51**])<br>1.3.4. Hackathons (John Gage and OpenBSD, 1999 [**54**])<br>1.3.5. Lean Startup Methodology (Eric Ries, 2011 [**59**]) |
| | 1.4. Computational and AI-Driven Methods | 1.4.1. Genetic algorithms and evolutionary computation (John Holland, 1975 [**60**])<br>1.4.2. Automated hypothesis generation (Don R. Swanson, 1986 [**61**])<br>1.4.3. Generative adversarial networks for idea synthesis (Ian Goodfellow, 2014 [**66**])<br>1.4.4. Large language models for ideation support (Ashish Vaswani et al., 2017 [**67**], OpenAI, 2019 [**68**]) |
| 2. IDEA EVALUATION METHODS (techniques to judge the quality, feasibility, novelty, or value of proposed ideas) | 2.1. Expert-Based Evaluation | 2.1.1. Peer review (Henry Oldenburg, 1665 [**70**]) <br> 2.1.2. Modified Delphi for scoring (Norman Dalkey and Olaf Helmer, RAND, 1963 [**34**]) <br> 2.1.3. Expert panels and consensus conferences (National Institutes of Health, 1977, 1990 [**72**]) <br> 2.1.4. Analytical Hierarchy Process (Thomas L. Saaty, 1977,1980 [**74,75**]) |
| | 2.2. Quantitative Assessment Metrics | 2.2.1. Cost-effectiveness and cost-benefit analysis (Abbé de Saint-Pierre, 1708 [**85**]; Ezra J. Mishan, 1971 [**88**]) <br> 2.2.2. Net present value (NPV) and internal rate of return (IRR) (Irving Fisher, 1907 [**91**]; Joel Dean, 1951 [**93**]) <br> 2.2.3. Patent metrics (US Patent & TM Office, 1947; Adam Jaffe, Manuel Trajtenberg, Bronwyn Hall, 1993 [**93**]) <br> 2.2.4. Bibliometric and scientometric indices (Eugene Garfield, 1955 [**96**] and 1972 [**97**]; Jorge Hirsch, 2005 [**98**]) <br> 2.2.5. Technology Readiness Levels (Stanley Sadin, John C. Mankins and NASA, 1970s, 1995 [**81**]) |
| | 2.3. Scoring Models and Criteria-Based Frameworks | 2.3.1. Multi-criteria decision analysis (MCDA) (Harold W. Kuhn and Albert W. Tucker, 1951 [**97**]) <br> 2.3.2. SWOT analysis - Strengths, Weaknesses, Opportunities, Threats (Albert S. Humphrey, 1960s [**105**]) <br> 2.3.3. Weighted scoring models (Stanley Zionts, 1979 [**101**]) <br> 2.3.4. Pugh matrix - decision-matrix method (Stuart Pugh, 1980s, 1990 [**110**]) <br> 2.3.5. Feasibility–Desirability–Viability framework (Tim Brown, 2009 [**50**]) |
| | 2.4. Crowd-Based Assessment | 2.4.1. Wisdom of the crowd techniques (Francis Galton, 1907 [**115**]) <br> 2.4.2. Prediction markets (Robin Hanson, 1980s, 1990 [**117**]) <br> 2.4.3. James Lind Alliance (Nick Partridge and John Scadding, 2004 [**40**], with Iain Chalmers) <br> 2.4.4. Child Health and Nutrition Research Initiative (CHNRI) method (Igor Rudan, 2006, 2008 [**41,42**]) <br> 2.4.5. Social media engagement metrics as proxies for idea traction (Jason Priem, 2010 [**127**]) |
| | 2.5. Scientific and Philosophical Validity Tests | 2.5.1. Logical consistency and deductive reasoning (Aristotle, 4th century BC [**133,134**]) <br> 2.5.3. Empirical testability, replicability and falsifiability (Francis Bacon, 1620 [**135**]; Karl Popper, 1934 [**139**]) <br> 2.5.3. Paradigm shift (Thomas Kuhn, 1962 [**140**]) |
| 3. IDEA PRIORITIZATION | 3.1. Structured Decision-Making Frameworks | 3.1.1. Paired comparison methods (Louis L. Thurstone, 1927 [**141**]) <br> 3.1.2. Multi-voting and Dot-voting (group facilitation practices, 1950s–1960s [**144,145**]) <br> 3.1.3. Delphi with ranking rounds (Norman Dalkey and Olaf Helmer, RAND, 1963 [**34**]) <br> 3.1.4. Nominal Group Technique with voting (Andre L. Delbecq and Andrew H. Van de Ven, 1971 [**35**]) <br> 3.1.5. Analytic Hierarchy Process (Thomas L. Saaty, 1977,1980 [**74,75**]) |

| METHODS (methods select the most promising ideas for action, investment, or further study) | 3.2. Priority-Setting Frameworks in Health and Science | 3.2.1. RAND/UCLA Appropriateness Method (RAND Corporation and UCLA clinicians, 1980s, 2001 [**146**]) |
| --- | --- | --- |
| | | 3.2.2. Essential National Health Research framework (COHRED, 1990 [**147**]) |
| | | 3.2.3. GRADE methodology with EtD (Evidence to Decision) frameworks (GRADE Working Group, 2000 [**149**]) |
| | | 3.2.4. James Lind Alliance Partnerships (Nick Partridge and John Scadding, 2004 [**40**], with Iain Chalmers) |
| | | 3.2.5. Combined Approach Matrix (Abdul Ghaffar, Andres de Francisco, Stephen Matlin, 2004; [**152**]) |
| | | 3.2.6. Child Health and Nutrition Research Initiative (CHNRI) method (Igor Rudan, 2006, 2008 [**41,42**]) |
| | 3.3. Portfolio and Pipeline Management Tools | 3.3.1. Real Options Analysis (Stewart C. Myers, 1947 [**156**]) |
| | | 3.3.2. R&D portfolio matrices (Bruce Henderson and Boston Consulting Group, 1970 [**159**]) |
| | | 3.3.3. Product roadmapping and prioritization grids (Robert Phaal and colleagues, 1970s to 2000s, 2004 [**162**]) |
| | | 3.3.4. Stage-gate model (Robert G. Cooper, 1980s, 1990 [**164**]) |
| | 3.4. AI-Driven Prioritization Tools | 3.4.1. Knowledge graphs and semantic similarity clustering (Allan M. Collins and M. Ross Quillian, 1960s [**165**]) |
| | | 3.4.2. Reinforcement learning-based portfolio optimization (Richard S. Sutton and Andrew G. Barto, 1998 [**179**]) |
| | | 3.4.3. Automated priority setting via large language models (Peige Song and Igor Rudan, ISoGH, 2024 [**155**]) |
| | 3.5. Participatory and Democratic Prioritization | 3.5.1. Cross-Cutting Philosophical and Meta-Theoretical Approaches (Paul Feyerband, 1975 [**188**], and others) |
| | | 3.5.2. Citizen juries and deliberative democracy forums (Ned Crosby, 1974 [**194**]; James Fishkin, 1991 [**196**]) |
| | | 3.5.3. Public consultations and e-surveys with weighting (Stephen Sedley, 1985 [**198**]) |
| | | 3.5.4. Participatory budgeting (Tarso Genro and Raul Pont, 1989 [**203-205**]) |
| | 3.6. Approaches to prioritising ideas beyond specific techniques | 3.6.1. Occam's Razor (William of Ockham, 1323-1328 [**208,209**]) |
| | | 3.6.2. Bayesian inference (Thomas Bayes, 1763 [**213**]) |
| | | 3.6.3. Dialectical method (Georg Wilhelm Friedrich Hegel, 1807 [**215**]) |
| | | 3.6.4. Epistemic humility and pluralism (John Stuart Mill, 1859 [**218**]) |

The philosophical foundations of this project were laid in three earlier conceptual papers by one of the co-authors (IR). In the first one, he proposed that the human brain might be understood not merely as a rational processor or associative network of information gathered by the senses, but as a sensor of ideas itself - an evolved organ that continuously perceives ideas or generates them, evaluates them, and then prioritises them based on the core criteria of attractiveness (i.e., a subjective, emotional component), feasibility (i.e., an objective, rational component), and potential future impact (i.e., a foresight based on an appropriate understanding and valuing of the information available about the context, which includes the brain's ability to model time, space and events) [**1**]. In the second paper, he then explored the value of information, a concept traditionally rooted in economics and game theory, as a foundational principle for evaluation of ideas within the context, along with the dangers of disinformation for perception of ideas [**7**]. He argued that information influences the sense of ideas through the dimensions of relevance, credibility and leverage, thus having capacity to reduce uncertainty, improve prediction, and inform future actions in a physical world. This logic applies equally to scientific hypotheses, business strategies, and ethical frameworks. In the third paper, he focused on what makes science successful, and what is necessary for a field of science to originate, thrive, and eventually start to fulfil its mission [**8**].

Building on those philosophical insights, this work takes a more practical turn: we attempt, to the best of our knowledge, the first comprehensive synthesis of the many methods that humans have developed and used across numerous disciplines to generate, evaluate, and prioritise ideas. We fully acknowledge the futility of trying to identify every tool, framework, or heuristic ever devised. The intellectual landscape is too vast, and many methods are proprietary, undocumented, or culturally specific. Also, historically, those in power often resorted to various forms of propaganda or violence to ensure that only their ideas thrived, while those of their opponents were censored and silenced [**9,10**]. We do not think of such approaches to prioritising ideas as scientifically structured or grounded, but we do address their clear impact in the discussion section. Nonetheless, by tracing the evolution of major western techniques, spanning from the classical heuristics of Aristotle and Bacon [**11**] to contemporary AI-powered systems, we aim to present the first map of the Ideometrics landscape, presenting it in a format that is amendable in the future.

Expectedly, most of these methods originated in the social sciences and humanities, often grounded in qualitative reasoning, normative deliberation, and participatory judgment. Techniques like the dialectical method, epistemic humility and pluralism, Delphi method, focus groups, citizen juries, public consultations, and cross-cutting philosophical and meta-theoretical approaches, were developed not merely to reach technical decisions, but to navigate value pluralism, stakeholder inclusion, and legitimacy in contested spaces. Others, like peer review and consensus conferences, aimed to institutionalise epistemic standards without reducing science to algorithmic scoring [**12**]. However, in recent decades, the landscape has been transformed by the advent of quantitative metrics, digital tools, and AI. Citation counts, h-indices, and journal impact factors introduced the notion that influence could be numerically tracked. Cost-benefit and cost-effectiveness analyses formalised trade-offs in economic and policy domains. Crowdsourcing platforms, real-time feedback systems, and online participatory tools allowed for idea generation and ranking at unprecedented scale and speed. Most recently, large language models (LLMs), knowledge graphs, and reinforcement learning algorithms have begun to automate parts of the idea evaluation and prioritisation pipeline, raising new questions about the future of human creativity, judgment, and responsibility [**13**].

We believe that this moment of convergence between traditional humanistic methods and emerging machine-assisted systems demands a structured understanding of the field. Proposing Ideometrics as a new scientific field is our attempt to provide such a structure. The framework we propose divides the field into three principal domains:

1. *Idea Generation*, encompassing classical creativity techniques, collaborative design practices, and computational ideation tools, from brainstorming and Theory of Inventive Problem Solving (TRIZ), CHNRI's expert crowdsourcing, to mind mapping, hackathons, and large language models (LLM);
2. *Idea Evaluation*, covering both expert-driven and data-driven methods, including peer review, Delphi panels, Analytic Hierarchy Process (AHP), citation and patent metrics, cost-effectiveness analysis, CHNRI collective scoring, and foundational principles like falsifiability and testability;
3. *Idea Prioritisation*, including structured decision tools such as the Stage-Gate model, CHNRI priority scores, Grading of Recommendations Assessment, Development and Evaluation (GRADE), portfolio matrices, as well as participatory and philosophical approaches like participatory budgeting, citizen deliberation, and epistemic humility.

Within each of these domains, we attempt to trace the historical origins, intellectual foundations, key contributors, methodological innovations, and current applications of each major approach. We also examine how different traditions overlap, diverge, or can be integrated, particularly as digital and AI-based methods gain prominence. Importantly, we view this work not as a static catalogue, but merely as the *first iteration of a living framework*. As new methods emerge, and the existing ones are unearthed or gain more visibility in the scientific literature, whether through technological innovation, social experimentation, or philosophical reflection, this structure should allow for their systematic incorporation and comparative analysis. We anticipate that as fields such as science policy, global health, digital governance, and AI ethics continue to grapple with how to prioritise among competing visions, the existing and emerging methodologies of Ideometrics will prove increasingly relevant.

In summary, this work represents an initial, imperfect, and incomplete, but necessary step toward unifying the study of ideas as structured phenomena, subject to scientific inquiry. We believe that by bringing together centuries of fragmented knowledge into a coherent whole, we can better equip individuals, organisations, and societies to navigate the abundance of new ideas in the 21st century, and to focus their attention on those most likely to have a positive impact on humanity's present and future.

## 1. Idea Generation Methods

### 1.1. Classic Approaches to Idea Generation: Individual and Cognitive Techniques

#### 1.1.1. Stream of Consciousness and Freewriting

Freewriting and stream-of-consciousness writing are two of the most enduring techniques for unlocking creativity. They share a commitment to spontaneity and unfiltered thought. Although often mentioned together, they emerged from distinct intellectual traditions and served different purposes. The philosophical foundations of stream-of-consciousness writing can be traced to the American psychologist and philosopher William James. In his landmark work in 1890, *The Principles of Psychology* [**14**], James introduced the concept of the "stream of thought," later popularised as the "stream of consciousness." He described consciousness as a flowing, continuous succession of ideas and sensations: an ever-shifting experience rather than a static sequence of mental events. His insights, especially the emphasis on non-linear and introspective awareness, proved profoundly influential in shaping 20th-century literature. Although James did not formally propose stream-of-consciousness writing as a literary technique, his concept of "consciousness as a continuous flow" inspired literary modernists such as James Joyce, Virginia Woolf, and William Faulkner, who adopted his idea to explore the complexities of human conscious thought in their novels. The style first appeared in Édouard Dujardin's 1887 novel *Les Lauriers sont coupés* [**15**], but reached broader recognition with James Joyce's *Ulysses* in 1922 [**16**]. These authors developed stream-of-consciousness into a literary device meant to simulate the internal monologue and fragmented experience of consciousness, rather than to generate ideas *per se*. Freewriting, by contrast, emerged in the 1970s as a pedagogical tool. Peter Elbow, in his influential 1973 book *Writing Without Teachers* [**17**], introduced freewriting as a method for overcoming internal censorship and accessing subconscious thought. The technique involves writing continuously for 10–15 minutes without editing, pausing, or planning. This allows thoughts to flow in language with minimal inhibition. While Elbow was inspired by earlier educational theories and perhaps indirectly by Jamesian introspection, he was the first to formalise freewriting as a systematic approach to idea generation in the classroom. The genealogy of these methods reflects a fascinating convergence: Jamesian psychology led to modernist literary experimentation, and then to Elbow's pedagogical formalization. While stream-of-consciousness writing aimed to depict the raw texture of thought, freewriting sought to liberate the creative process itself.

### 1.1.2. Theory of Inventive Problem Solving (TRIZ)

TRIZ, the Russian acronym for "Theory of Inventive Problem Solving," was developed by Soviet navy engineer and patent examiner Genrich S. Altshuller. Beginning his work in 1946, Altshuller analysed thousands of patents in search of recurring patterns in inventive solutions. This analysis led to the formulation of a structured methodology for creativity in engineering and innovation. Although imprisoned during Stalin's purges from 1950 to 1954, Altshuller resumed his research upon release. He introduced TRIZ formally in 1956 [**18**]. He continued to refine TRIZ over the next several decades. His major English-language work, *Creativity as an Exact Science* in 1984, outlined the core components of TRIZ, including dozens of inventive principles, engineering parameters, the contradiction matrix, and patterns of technological evolution [**19**]. TRIZ provides a systematic approach to overcoming technical contradictions without compromise. By abstracting problems and applying universal solution patterns derived from successful inventions, TRIZ seeks to enable innovation through structured logic rather than trial and error.

### 1.1.3. Morphological Analysis

Morphological analysis is a method of generating creative solutions by systematically exploring all permutations of relevant variables in a structured framework. It was introduced by Swiss astrophysicist working at Caltech, Fritz Zwicky, during the 1940s and 1950s. It was an effort to make invention more routine within aeronautics. Zwicky's work culminated in two landmark publications: *Morphological Astronomy* in 1957 [**20**] and *Discovery, Invention, Research – Through the Morphological Approach* in 1969 [**21**]. His method involved identifying key parameters of a problem, listing all possible values or states for each parameter, creating a multi-dimensional matrix (a "Zwicky box"), and exploring combinations to uncover novel, feasible configurations. It exhaustively maps the solution space and the landscape of potential configurations to encourage idea generation through recombination, thus revealing previously overlooked solutions. It has proven especially useful in complex problem structuring, scenario planning, and exploratory design.

### 1.1.4. Lateral Thinking

Lateral thinking, a term that entered popular culture for deliberate creativity, was coined in 1967 by Edward de Bono, a Maltese physician, psychologist, and creativity theorist and Oxford University's Rhodes scholar. Unlike traditional "vertical" thinking, which proceeds logically and sequentially, lateral thinking seeks to disrupt habitual mental patterns to provoke fresh insights. De Bono introduced this concept in his book *The Use of Lateral Thinking* in 1967 [**22**]. He outlined several practical techniques, including breaking out of dominant mental patterns, using deliberate techniques to generate new perspectives and ideas, encouraging discontinuous thinking rather than sequential logic, and challenging assumptions and established frames. Unlike brainstorming or logical deduction, which are discussed later in the text, lateral thinking provokes unexpected connections and reversals to enable creative insight, escaping conventional logic and discovering unexpected solutions. By breaking assumptions and encouraging discontinuities in thought, lateral thinking formalised the idea that creativity could be taught and systematised. It led to innovative training programs in education, strategy, and organisational problem-solving.

### 1.1.5. SCAMPER Technique

The SCAMPER technique is a structured ideation tool developed by Robert F. Eberle, an American creativity educator and promoter, and introduced in his book *SCAMPER: Games for Imagination Development* in 1971 [**23**]. Originally aimed at helping children learn how to think creatively, it was quickly adopted in business and product development for its practical utility. SCAMPER is an acronym that stands for: Substitute - Combine – Adapt - Modify (or Magnify) - Put to another use – Eliminate - Reverse (or Rearrange). Each prompt encourages the user to manipulate an existing idea in a specific way, enabling systematic variation and creative expansion. SCAMPER evolved from earlier brainstorming methods introduced by Alex Osborn (see heading 1.2.3.), but stands out for its operational simplicity and memorable structure.

### 1.1.6. Mind Mapping

Mind mapping was introduced by British psychologist and author Tony Buzan in the early 1970s. It gained broader attention through his BBC television series *Use Your Head* [**24**], and was further developed in *The Mind Map Book* in 1993 [**25**]. Mind maps begin with a central idea and radiate outward in branches, using keywords, images, colours, and spatial layout to mirror the brain's associative thinking. This non-linear format facilitates ideation, improves memory, and enhances understanding by engaging both hemispheres of the brain. Mind mapping is particularly useful in brainstorming, planning, studying, and visual problem-solving. Its popularity stems from its intuitive format and cognitive alignment with how people naturally organize information.

### 1.1.7. Heuristics and Biases Framework

Although not designed specifically for ideation, the heuristics and biases framework, developed by Israeli psychologists Daniel Kahneman and Amos Tversky, working together at the Hebrew University in Jerusalem, has significantly influenced how we understand idea formation and evaluation. First articulated in their 1974 *Science* article "Judgment Under Uncertainty: Heuristics and Biases" [**26**], the framework identified mental shortcuts such as representativeness, availability, and anchoring. While these heuristics facilitate fast decision-making, they often lead to systematic cognitive errors or biases. This body of work helped explain why some ideas gain traction and others are prematurely dismissed. It laid the foundation for debiasing strategies, such as "devil's advocacy", red teaming, and structured decision protocols, that now enhance ideation processes by counteracting intuitive blind spots. The collaboration between Kahneman and Twersky is credited to have established behavioural economics; Kahneman later moved to Princeton University, USA, and received the Nobel Prize in Economics in 2002 for integrating insights from psychological research into economic science, especially concerning human judgment and decision-making under uncertainty. His lifetime work has been summarised in a popular book *Thinking, Fast and Slow* in 2011 [**27**].

### 1.1.8. Six Thinking Hats

Also developed by Edward de Bono, the *Six Thinking Hats* technique was introduced in the 1985 book [**28**] to structure group thinking and ideation. Each "hat" represents a distinct cognitive role or mode: *White*: Neutral facts and data; *Red*: Emotions and intuition; *Black*: Caution and critical judgment; *Yellow*: Optimism and benefits; *Green*: Creativity and alternatives; *Blue*: Process and meta-thinking. By synchronizing group attention on one thinking mode at a time, this method reduces conflict, enhances focus, and promotes constructive ideation. The Green Hat is explicitly used for creative generation, but all hats contribute to well-rounded idea development.

## 1.2. Participatory, Collaborative Methods and Surveys for Idea Generation: Group-Based and Social Approaches

### 1.2.1. Brainstorming

Brainstorming was introduced by advertising executive Alex F. Osborn in the late 1930s to address the lack of innovation in team settings at his advertising agency Batten, Barton,

Durstine & Osborn (BBDO). He first described the technique briefly in his book *How to "Think Up"* in 1942 [**29**] and expanded it fully in *Applied Imagination* in 1953 [**30**]. Osborn's method emphasised four cardinal rules: defer judgment, strive for quantity, encourage wild ideas, and combine and improve ideas, while building on the ideas of others. These principles were intended to foster psychological safety, stimulate divergent thinking and creative collaboration, and break habitual patterns of thought in group settings. Despite later critiques questioning its comparative effectiveness in groups versus individuals, brainstorming laid the foundation for later creativity frameworks including SCAMPER, TRIZ, lateral thinking and "design thinking". It remains a widely used starting point for ideation in education, business, and research.

### 1.2.2. Focus Groups and In-Depth Interviews

Group interviewing techniques, a precursor to focus groups, were used in the 1920s for developing survey questionnaires and other social research purposes. Focussed interviews emerged as a systematic method for eliciting group insights in the early 1940s, introduced by sociologist Robert K. Merton in collaboration with Patricia L. Kendall and Marjorie Fiske. Working for the U.S. Office of War Information during World War II, Merton sought to understand audience responses to radio broadcasts and propaganda through what he termed "focused interviews." These interviews emphasized participant interaction, spontaneous discussion, and subjective reactions to specific stimuli. The foundational publication, *The Focused Interview: A Manual of Problems and Procedures* (1956) codified this method [**31**]. It introduced the principles of non-directive moderation, guided but open-ended questioning, and the analytical value of group dynamics. Although the term "focus group" gained popularity and recognition later, popularised by Ernest Dichter [**32**] and other scholars, Merton's methodological contribution laid the groundwork for its widespread adoption. Initially rooted in social research, focus groups expanded into market research, healthcare, public policy, and innovation by the latter half of the 20th century. Their ability to generate and explore perceptions, emotions, and preferences made them especially useful in idea generation, concept testing, co-creation workshops, and participatory design. By the 1980s, works such as David L. Morgan's *Focus Groups as Qualitative Research* helped formalize the approach in academic literature [**33**]. Today, focus groups remain a cornerstone of qualitative inquiry and idea generation across diverse fields, valued for their capacity to surface detailed and unexpected insights through structured yet flexible dialogue.

### 1.2.3. Delphi Technique

The Delphi method was developed in the 1950s by Norman Dalkey and Olaf Helmer at the RAND Corporation as a tool for structured expert consultation, initially aimed at military and technological forecasting during the Cold War. Its foundational principles were iterative surveys, anonymity, controlled feedback, and statistical synthesis. They were all designed to minimise bias from group dynamics while enabling "collective intelligence". The technique was first formally introduced to the academic community in the paper *An Experimental Application of the Delphi Method to the Use of Experts* In 1963, where its procedures and logic were outlined [**34**]. Experts respond to a series of questionnaires in multiple rounds, with anonymity to reduce dominance bias. Controlled feedback is provided after each round, with responses aggregated and refined after each round until convergence towards consensus is achieved. Though originally intended for forecasting, the Delphi method quickly found applications in strategic planning, health research prioritization, foresight exercises, and multidisciplinary consensus-building [**34**]. Its rigorous yet flexible structure makes it especially suitable for contexts where empirical data is limited, and expert judgment is essential.

### 1.2.4. Nominal Group Technique (NGT)

The Nominal Group Technique was introduced in the late 1960s by organisational theorists Andre L. Delbecq, Andrew H. Van de Ven, and later David H. Gustafson at the University of Wisconsin–Madison. NGT aims to balance individual ideation with structured group decision-making. It is a structured approach that encourages equal participation and minimises conformity pressure, the influence of dominant personalities or group dynamics on idea generation, making it especially useful for diverse groups. Participants first generate ideas silently and independently. Then, in a round-robin format, they share ideas without discussion. After clarification, they privately vote or rank ideas to establish group priorities. This structured flow ensures balanced participation and clear prioritization. The technique was described in their article *A Group Process Model for Problem Identification and Program Planning* in 1971 [**35**], with a definitive guide and distinction from brainstorming and Delphi processes in the book *Group Techniques for Program Planning* in 1975 [**36**]. NGT has since become a popular tool in public health, strategic planning, and participatory research settings.

### 1.2.5. World Café

The World Café method was introduced by Juanita Brown and David Isaacs in 1995. They were designing strategic dialogue forums with senior leaders. They came up with the World Café as a way to harness collective intelligence through structured conversation. Inspired by an impromptu café-style dialogue at real-life meetings, they formalised the method into a participatory approach for large-scale idea generation. The process simulates a café environment with small tables, rotating participants, and a "table host" who captures the evolving dialogue. Over successive rounds, ideas are cross-pollinated, deepened, and synthesized. A concluding "harvest" session captures emerging themes. Their 2005 book, *The World Café: Shaping our futures through conversations that matter*, details the principles of the design, offers implementation guidance, and showcases real-world applications in education, business, and public policy [**37**].

### 1.2.6. Open Space Technology

Open Space Technology (OST) was introduced in the mid-1980s by Harrison Owen, an Episcopal priest whose academic training focused on the nature and function of myth, ritual, and culture. It is a self-organising facilitation method for addressing complex issues. Inspired by the informal creativity observed during conference coffee breaks, Owen created a format in which participants build their own agendas around a central theme. Key features include an open marketplace of ideas, voluntary participation, and flexible session structure guided by principles such as: "Whoever comes is the right people" and "The law of two feet"

(encouraging people to move between sessions based on interest). OST was first implemented in 1985 and formally described in Owen's book *Open Space Technology: A User's Guide* in 1997 [**38**]. It has since been adopted in change management, crisis response, multi-stakeholder engagement, and innovation workshops.

### 1.2.7. InnoCentive and other "open innovation" platforms

InnoCentive, a platform for "open innovation", was conceived by Alpheus Bingham and Aaron Schacht in 1998 at the Eli Lilly pharmaceutical company. The initial idea evolved into "Molecule.com" and then "BountyChem" before becoming InnoCentive. Bingham, then a senior executive at Eli Lilly, helped create InnoCentive as an external innovation platform to tackle research and development bottlenecks. The company was launched in 2001 and became a spin-out in 2005. It sought to crowdsource solutions to challenges in research and development by connecting organizations with global problem-solvers. The platform invites "seekers" to post technical challenges, which are tackled by a distributed network of "solvers." Winning solutions receive cash prizes. The approach was elaborated in the book *The Open Innovation Marketplace* in 2011, co-authored by Alpheus Bingham and Dwayne Spradlin [**39**]. The book explains the logic of challenge-driven innovation (CDI). InnoCentive helped catalyze the broader open innovation movement, inspiring similar platforms in data science (Kaggle), global development (XPRIZE) and space technology (NASA's Centrifuge).

### 1.2.8. The James Lind Alliance (JLA) Priority Setting Partnerships (PSPs)

The James Lind Alliance (JLA) was initiated in 2004 by Sir Iain Chalmers, a British physician, health services researcher and one of the founders of the Cochrane Collaboration, with support from the UK Medical Research Council (MRC) and the National Institute for Health Research (NIHR). It was first described by its co-founders Sir Nick Partridge and John Scadding in the Lancet article in 2004 [**40**]. The goal of this non-profit initiative is to promote equitable and transparent research prioritisation by bringing together patients, carers, and clinicians in structured collaborations. These Priority Setting Partnerships (PSPs) aim to identify and rank unanswered questions in healthcare that are most relevant to those directly affected. The initiative was named after James Lind, the 18th-century Scottish naval surgeon credited with conducting one of the first controlled clinical trials on scurvy. The PSPs perform a series of steps: collection of uncertainties, verification against existing research, stakeholder surveys, and consensus workshops using voting-based techniques, often adapting the Nominal Group Technique. Final outcomes are prioritised "Top 10" lists that inform research funding, guideline development, and policy planning. JLA PSPs generate research ideas in the first step through a broad stakeholder survey, often referred to as the open call for the "initial survey of uncertainties." Then, patients, carers, and clinicians submit questions or uncertainties about a particular health condition or topic that they believe should be answered by research. Ideas are collected through an online or paper-based survey disseminated via charities, professional bodies, clinics, social media, etc. The survey is intentionally open-ended and respondents are not prompted with predefined questions. The method ensures that research ideas come directly from those most affected, and not from academics or policymakers, helping to surface real-world uncertainties that may not be apparent to researchers alone. The raw responses are then reviewed and interpreted by the PSP team, duplicates merged, out-of-scope entries removed, and submissions rewritten into researchable summary questions using the Patient/Population, Intervention, Comparison, and Outcome (PICO) format where possible. Each summary question is checked against existing evidence, such as systematic reviews, to ensure it is a true uncertainty. This bottom-up ideation approach ensures the prioritised research agenda is relevant, meaningful, and co-produced by those who live with or treat the condition.

**1.2.9. The Child Health and Nutrition Research Initiative (CHNRI) method**

The Child Health and Nutrition Research Initiative (CHNRI) method was initially developed in 2006 by Igor Rudan from the University of Edinburgh, a physician with interests in anthropology, human genetics, and global health economics. He was contracted as the consultant of the CHNRI of the Global Forum for Health Research in Geneva, Switzerland, by Professors Shams El Arifeen and Robert E. Black, who co-authored the first version of the CHNRI methodology's conceptual framework [**41**]. Detailed guidelines for implementation were then published by Rudan and colleagues in 2008 [**42**]. Originally focused on child health research in low- and middle-income countries, the method is credited for achieving several conceptual advances. They include generating hundreds of research ideas in a systematic and transparent way. Proposed by many leading experts in the field, the research ideas are grouped into: (i) those that can better describe the problem; (ii) those that can optimise the way in which current resources and interventions are addressing the problem; (iii) those that can further develop and improve the available interventions, to make them more cost-effective and/or equitable; and (iv) those that could discover entirely novel interventions. When applied to the field of health research, this "four D" framework – "description, delivery, development and discovery" – translated into epidemiological research, health systems and policy research, research on improving the existing health interventions, and research to discover new health interventions. From those four fundamental instruments of health research, hundreds of proposed research ideas would then be categorised into broad "research avenues", narrower "research options", and specific "research questions", accounting for the "depth and breadth" of each proposed idea. This method also enables analysis of the level of saturation of the proposed spectrum of ideas: it can trace the increase in probability that any further ideas would have already been suggested previously by contributing experts. Since 2008, the CHNRI method has been widely adopted and applied by the World Health Organisation (WHO), United Nations Children's Fund (UNICEF), national governments, academic collaborations and major global funders across diverse domains of health and development research. Its conceptual advances make it flexible and applicable to almost any problem in any area of science. Its approach to evaluating ideas and prioritising ideas will be reviewed in further headings.

**1.2.10. IdeaScale**

IdeaScale, launched in 2008-2009 by Vivek Bhaskaran and Rob Hoehn, is a web-based platform designed to crowdsource ideas and manage innovation workflows [**43,44**]. It is a cloud-based software company that licenses innovation management software employing crowdsourcing. The company was founded by Vivek Bhaskaran and Rob Hoehn in Seattle. It gained immediate traction when 23 U.S. federal agencies adopted it under President Obama's Open Government Initiative. The platform enables users to submit, vote on, comment, and refine ideas in a structured interface. A configurable workflow guides submissions from inception to evaluation and implementation, promoting collective intelligence in both public and corporate settings. While lacking a single key academic publication, IdeaScale's success is documented in industry case studies and civic tech literature. It remains a leading software-as-a-service (SaaS) solution for participatory innovation in innovation management.

**1.3. Generating Ideas for Creation: Design and Innovation Frameworks**

**1.3.1. Human-Centered Design**

Human-Centered Design (HCD) is an approach to innovation that begins with understanding the needs, behaviours, and experiences of the people for whom solutions are being developed. While the roots of the concept extend across disciplines, including engineering, psychology, anthropology, and the arts, the formalization of HCD began in the mid-20th century. One of the earliest proponents of the approach was Professor John E. Arnold, who taught creative engineering at MIT and later at Stanford, where he helped establish the design program in 1958 [**45**]. Arnold's work emphasized that engineering and product development should center on human needs, setting the stage for the philosophy that would later mature into HCD. The concept evolved significantly in the late 20th century, gaining traction in design, technology, and public service. Cognitive scientist Donald A. Norman played an important role in establishing the psychological foundations of user-centered design, most notably through his book *The Psychology of Everyday Things* in 1988, which highlighted usability, intuitive interfaces, and feedback systems [**46**]. HCD was further popularized by the design firm IDEO, founded by David Kelley, which helped institutionalize the methodology in business and development sectors. The key elements of the process are inspiration, through deeply understanding people's lives and challenges; then, ideation, through generating and refining ideas collaboratively; and implementation, through prototyping, testing, and iterating based on real-world feedback. The collaboration between IDEO and Stanford's d.school - officially Hasso Plattner Institute of Design - catalyzed a global movement in human-centered innovation. A widely used field manual, *The Field Guide to Human-Centered Design* by IDEO.org in 2015 [**47**], later made the methodology accessible to social innovators, NGOs, and policy practitioners.

**1.3.2. Design Thinking**

Design Thinking is a human-centered, iterative process for creative problem-solving, distinguished by its blend of analytical rigor and empathy-driven insight. Its intellectual roots trace back to Herbert Simon's *The Sciences of the Artificial* in 1969 [**48**], where he described design as a generic cognitive strategy for addressing ill-structured problems. Robert McKim's *Experiences in Visual Thinking* in 1973 [**49**] further emphasized creativity and visual reasoning in engineering design. But it was in the 1990s and 2000s that the method was crystallized and disseminated globally, primarily through the work of David Kelley, founder of IDEO and Stanford's d.school. IDEO's CEO Tim Brown articulated the philosophy in his influential book *Change by Design* in 2009 [**50**] and popularised the method. His book helped formalize Design Thinking as a structured method involving five non-linear stages: empathize, i.e., deeply understand users' needs; then, define, by framing the core problem; follow by ideating, through brainstorming creative solutions; develop a prototype, thus building tangible representations; and, finally, test, gather feedback and refine. The process emphasizes human-centeredness, rapid prototyping and iteration, and cross-disciplinary collaboration. Design Thinking has been widely adopted across industries, from business and education to healthcare and international development, often overlapping with, or incorporating Human-Centered Design.

### 1.3.3. Agile Ideation Sprints

Agile ideation sprints are short, structured cycles for collaboratively generating and validating ideas. They emerged from the Agile software development movement, particularly Scrum methodology, and were later formalized for design and innovation contexts by Google Ventures in the form of the Design Sprint. In the Agile movement, Jeff Sutherland and Ken Schwaber formalized Scrum in 1995, emphasizing iterative progress, team collaboration, and flexibility [51]. Although early sprints focused on delivering working software, they also fostered idea evolution during planning and review sessions. Manifesto for Agile Software Development, presented in 2001, introduced principles such as rapid iteration, collaboration, and working solutions over documentation [52] The concept of a dedicated "design sprint" was later introduced by Jake Knapp at Google Ventures in the early 2010s. Documented in the book *Sprint* in 2016, this five-day process guides teams from problem framing to tested prototype [53]. It combines principles from design thinking, user research, and agile workflows, making it accessible beyond software teams. The typical sprint includes mapping the problem, sketching ideas, deciding on a solution, prototyping, and testing with real users—all in a compressed timeframe [53].

### 1.3.4. Hackathons

Hackathons are intense, time-bounded events where individuals or teams collaborate to generate, develop, and prototype new ideas - often in the domains of software, hardware, or policy innovation. The term "hackathon" originates from the words "hack" and "marathon", where "hack" is used in the sense of exploratory programming, not a reference to breaching digital security [54]. It was independently coined in June 1999 during two separate events. The first was a cryptographic coding sprint organized by Theo de Raadt and the OpenBSD community in Calgary, Canada. Soon after, Sun Microsystems hosted the JavaOne Hackathon, where developers worked for 24 hours to build Java applications, marking the first branded corporate hackathon [54]. Initially focused on software engineering, hackathons quickly spread to fields like health innovation, civic technology, education, and sustainability. Their hallmark is rapid problem-solving under tight constraints, culminating in demo pitches to peers or judges. While no single publication introduced hackathons academically, key studies have since analyzed them as engines of innovation. Gerard Briscoe and Catherine Mulligan's paper *Digital Innovation: The Hackathon Phenomenon* in 2014 explored their role in digital ecosystems [55], Lilly Irani examined how they shape entrepreneurial culture in 2015 [56], while Thomas James Lodato and Carl DiSalvo studied them as vehicles for participatory and issue-driven design in 2016 [57].

### 1.3.5. Lean Startup Methodology

The Lean Startup methodology provides a systematic framework for developing ideas and innovations through iterative testing, rapid prototyping, and validated learning. Introduced by entrepreneur Eric Ries in the late 2000s, it draws from lean manufacturing, agile software development, and Steve Blank's customer development model [58]. Ries began promoting the framework through blog posts and talks around 2008, eventually formalizing it in his bestselling book *The Lean Startup* in 2011 [59]. The core cycle of *Build–Measure–Learn* encourages innovators to test hypotheses via Minimum Viable Products (MVPs), then launch quickly, measure performance, and learn from real user behavior, use learning to pivot or persevere, validate learning, measure progress by evidence that an idea works in the real world, track meaningful learning and progress through metrics tied to business hypotheses, and make structured course corrections based on feedback and data. It challenges traditional notions of success based on features or funding, instead measuring progress through evidence of what works. Concepts like "innovation accounting" and "pivoting" became part of the startup and corporate lexicon, with major adoption by companies like General Electric, Intuit, and government innovation programs.

**1.4. Approaches to Ideation in Digital Age: Computational and AI-Driven Methods**

**1.4.1. Genetic Algorithms and Evolutionary Computation**

Genetic Algorithms (GAs) and the broader domain of evolutionary computation apply principles of biological evolution to computational problem-solving and idea generation. This field was pioneered by John H. Holland, Professor of Psychology, Electrical Engineering, and Computer Science at the University of Michigan in the 1960s. He initially developed these concepts while studying cellular automata, and then tried to apply principles of biological evolution to computer systems. Holland formalized the theoretical foundations of GAs in his landmark work *Adaptation in Natural and Artificial Systems* in 1975 [60]. He encoded potential solutions as "chromosomes"/"strings", and then used mechanisms like crossover and mutation to evolve and optimise them over successive "generations." These solutions are evaluated using a fitness function, and refined through selection, crossover, and mutation, closely mimicking Darwinian processes. Key components of GAs include: *Chromosomes*: Encoded representations of candidate solutions; *Fitness Function*: Evaluates the performance of each solution; *Selection*: Favors better-performing solutions for reproduction; *Crossover*: Combines parts of parent solutions to form offspring; *Mutation*: Introduces variation to explore new possibilities; *Generations:* Repeat cycles of selection, recombination, and mutation over time. GAs have become a foundational tool in evolutionary computation, used across optimization, AI, design, and even computational creativity. Their applications include engineering design, logistics, neural architecture search in deep learning, robotics and game AI, and artistic generation.

**1.4.2. Automated Hypothesis Generation**

The concept of Automated Hypothesis Generation (AHG), where machines identify novel, testable scientific ideas, originated in the 1980s. Since then, it evolved into a powerful class of tools for discovery. The foundational idea was articulated by Don R. Swanson, a medical informatician at the University of Chicago, in 1986 through his "literature-based discovery" and the "theory of undiscovered public knowledge" [61]. Swanson demonstrated that connections could be inferred between disjoint literatures, for example, linking fish oil and Raynaud's syndrome via blood viscosity, by identifying intermediate "B-terms." This logic became the basis for platforms such as Arrowsmith, co-developed by Swanson and Neil Smalheiser [62], and later tools like LION LBD [63], IBM Watson Discovery [64], and Semantic Scholar [65]. These tools automate the identification of "B-terms" linking disconnected "A" and "C" domains (e.g., A → B → C). Based on this, they propose novel hypotheses, particularly in biomedicine and materials science. Recent advances leverage AI and graph-based embedding to generate hypotheses in neuroscience, physics, and chemistry.

### 1.4.3. Generative Adversarial Networks (GANs) for Idea Synthesis

Generative Adversarial Networks (GANs) were introduced by Ian Goodfellow, an American computer scientist, engineer and executive, a pioneer of artificial neural networks and deep learning. He worked as a research scientist at Google DeepMind and Google Brain, was the director of machine learning at Apple and one of the first employees at OpenAI. He developed generative adversarial network (GAN) in 2014 with his colleagues at the Université de Montréal [66]. This work revolutionized the landscape of machine creativity. GANs consist of two neural networks, a "generator" and a "discriminator", that compete in a zero-sum game. The generator produces synthetic outputs, while the discriminator attempts to distinguish real from fake data. Through this adversarial training process, the generator learns to create outputs that are increasingly indistinguishable from real data. This architecture enables the synthesis of realistic and novel content across a range of domains, including art, design, language, science, and drug discovery. It laid the groundwork for a vast field of research and application, with a large potential that is difficult to fully grasp at this time.

### 1.4.4. Large Language Models (LLMs) for Ideation Support

Large Language Models (LLMs) have become powerful tools for supporting ideation through natural language. Built upon the transformer architecture introduced by Ashish Vaswani and colleagues in 2017 [67], LLMs evolved from statistical language models into general-purpose reasoning and creativity engines. OpenAI's GPT series was the first to demonstrate emergent ideation capabilities at scale. GPT-2 revealed early signs of generalization, creativity and ideation abilities with zero-shot and few-shot learning in 2019 [68]. Soon thereafter, GPT-3 introduced robust few-shot learning, reasoning, synthesis, and creative, domain-crossing ideation in 2020 [69]. These models could generate hypotheses, expand on prompts, synthesize analogies, and co-create with human users in diverse fields, becoming "brainstorming partners" in fiction, design, business, science, policy and other fields. LLMs learned these capabilities not through hardcoded rules, but through massive-scale pretraining on internet-scale corpora, enabling them to recombine concepts in novel ways. Therefore, LLMs learned to "ideate" not by explicit programming, but through unsupervised pretraining on the entire internet, enabling latent concept combination and abstraction. Therefore, LLMs now routinely assist in research ideation, scientific writing, product design, fiction, strategy, and many other tasks, constituting a new era of "machine-augmented creativity". As of 2024, models like GPT-4, Claude, Gemini, DeepSeek, and open-source systems such as LLaMA and Mistral continue to push the frontiers of AI-supported creativity – again, with seemingly vast potential that is difficult to predict at this time.

# 2. Idea Evaluation Methods

## 2.1. Classical Methods for Evaluating Ideas: Expert-Based Evaluation

### 2.1.1. Peer Review

The practice of peer review, which consists of soliciting judgments from experts in subject matter to assess the validity, originality, and relevance of research ideas, has a long but gradual history. While informal scholarly critique dates back centuries, the structured, editorial peer review system as practiced today is a relatively modern institutional development. The earliest known examples of peer-like evaluation emerged in the 17th century, particularly with the Royal Society of London and its publication *Philosophical Transactions*, edited by Henry Oldenburg who introduced it in 1665 [**70**]. In France, a similar practice has been associated with Journal des Sçavans. Oldenburg is often regarded as a precursor of the peer review model due to his habit of seeking informal feedback on submissions from other scholars. However, this process was inconsistent, highly discretionary, and lacked transparency. Throughout the 19th century, prominent journals such as *The Lancet* and *Nature* relied largely on editorial judgment rather than external review. The contemporary model of anonymous or named external experts rigorously evaluating submitted manuscripts only emerged in the mid-20th century. This transformation was catalysed by several post–World War II developments, including a surge in scientific output and the expansion of publicly funded research via institutions such as the National Institutes of Health (NIH) and the National Science Foundation (NSF). These trends created an urgent need for systematic, fair, and expert-driven evaluation mechanisms to allocate limited publication space and funding. One of the earliest pioneers in formalizing peer review was the Journal of the American Medical Association (JAMA), which began implementing structured review processes in the 1940s. However, even leading journals like the New England Journal of Medicine (NEJM) did not adopt external peer review until 1976. A landmark analysis of the historical development of editorial peer review was provided by John C. Burnham in JAMA in 1990 [**71**]. This paper traces how editorial judgment gave way to systematic external scrutiny, documenting the social and institutional forces that shaped the peer review model as a gatekeeping mechanism in modern science.

### 2.1.2. Modified Delphi for Scoring

The origins of the Delphi method at the RAND corporation were described in the previous chapter. [**34**]. It was a helpful structured technique for eliciting expert judgment, particularly well-suited to situations with incomplete data, high uncertainty, or multi-disciplinary inputs. Initially, it was a part of classified military forecasting projects during the Cold War, so although it was used since 1950s, it became publicly known in 1963 [**34**]. The key innovation was a systematic approach to achieving expert consensus through multiple rounds of anonymous input, controlled feedback, and statistical aggregation of opinions. In this subheading, we reflect on the way that it evaluated proposed ideas. This was done through iterative rounds of input with controlled feedback, and then statistical convergence toward consensus and quantifiable scoring of subjective judgments. The Delphi method uses median or mean scores for any quantitative items, supporting it by frequency distributions or histograms to show variability among the scorers and thematic analysis for any qualitative comments. This aggregation represents the collective judgment of the panel in the first round, which is then followed by feedback to experts, showing how their responses compare to the group's central tendency, how much disagreement or convergence exists, and any justifications or rationales offered by other participants. This step allows participants to re-evaluate their own positions in light of the group's thinking. Then, experts are asked to reassess their initial responses and re-rate the ideas across two to four rounds, typically using either Likert scales (e.g., 1–9 for importance, feasibility, and impact), or ranking or scoring of competing ideas. Over successive rounds, the aim is to move toward consensus, though some level of dissent is acceptable and even encouraged to identify uncertainty or alternative scenarios. By the final round, ideas that have achieved high median ratings, narrow interquartile ranges that indicate consensus, and stable ratings across rounds are considered robust, high-priority, or widely supported. Conversely, ideas with low scores or wide disagreement may be deprioritized or marked for further study. Therefore, evaluation of ideas in Delphi process is, in essence, both quantitative (through aggregated expert ratings, medians, and convergence indicators), qualitative (through open comments, rationales, and recorded justifications), iterative (refining judgments through exposure to anonymous peer perspectives), and systematic (using structured questionnaires and statistical summaries). Therefore, although it was originally designed for forecasting technological developments and national security scenarios, the Delphi method was quickly adapted for use in healthcare, education, and research prioritization, particularly where empirical data were limited or absent. Its core strength lied in its ability to combine subjective expert opinions and insights with procedural rigor and replicability [**34**].

### 2.1.3. Expert Panels and Consensus Conferences

Structured expert panels and consensus conferences represent another classical approach to evaluating and prioritizing research ideas, particularly in contexts where high-stakes decisions must be made in the absence of definitive empirical evidence. These methods gained prominence during the 1970s and 1980s, especially in the field of health policy. The NIH Consensus Development Program, launched in 1977, institutionalized the practice of Consensus Development Conferences (CDCs) [**72**]. These brought together multidisciplinary panels of experts to review the available evidence, hear stakeholder input, and arrive at public consensus statements on key medical or research issues. The Delphi technique and other consensus development techniques, such as the NGT, differed in their structure and anonymity from CDCs, but they all aimed to facilitate reliable judgments among diverse experts. These methods became standard in organizations such as the NIH, the WHO, Agency for Healthcare Research and Quality (AHRQ), and various national research councils. Much of the foundational documentation on consensus conferences resides in institutional reports, highlighting the role of expert panel methods in transforming subjective expertise into defensible, consensus-based research recommendations [**73**].

### 2.1.4. Analytical Hierarchy Process (AHP)

The Analytical Hierarchy Process (AHP) is a mathematical and psychological framework for structured decision-making under complex, multi-criteria conditions. Developed by Thomas L. Saaty, a professor at the Joseph M. Katz Graduate School of Business at the University of

Pittsburgh in the 1970s, AHP allows for transparent prioritization by decomposing decisions into hierarchical elements and comparing them pairwise [74]. The key components of the method's contribution to evaluating ideas are decomposition of complex problems into hierarchical levels, their pairwise comparisons using a scale from 1 to 9, weighting of alternatives using eigenvalue computation and consistency ratio to assess reliability of judgments. AHP was further elaborated in Saaty's book *The Analytic Hierarchy Process: Planning, Priority Setting, Resource Allocation* in 1980 [75]. Originally applied to engineering and military strategy, AHP was soon adopted in public policy, health research, and strategic planning. Its ability to make trade-offs explicit made it particularly attractive for research funding allocation, technology assessment, and national science agenda setting, with adoption by institutions such as the National Science Foundation and the European Commission's Framework Programmes. AHP became one of the first structured, quantitative tools applied to evaluate and prioritize research ideas based on multiple weighted criteria, such as scientific merit, feasibility, societal impact and cost-effectiveness. Its quantitative approach and transparency allowed it to be combined with Delphi panels or newer frameworks like the CHNRI method, especially when multiple stakeholders must weigh diverse criteria.

## 2.2. Crunching the Numbers: Quantitative Assessment Metrics

### 2.2.1. Cost-Effectiveness and Cost-Benefit Analysis

Two additional tools that have profoundly influenced how ideas and projects are evaluated, particularly in public policy and health economics, are Cost-Benefit Analysis (CBA) and Cost-Effectiveness Analysis (CEA). They share roots in welfare economics, but were formalized in distinct institutional and disciplinary contexts. The roots of CBA can be traced back to early economic thinkers like Abbé de Saint-Pierre in France in 1708, who conducted one of the earliest cost-benefit analyses, specifically focusing on the utility of road improvements. In 19th-century France, Jules Dupuit introduced the concept of consumer's surplus that established the economic basis of CBA [76]. In 1890 Alfred Marshall, a prominent British economist, focused on supply and demand and publishes his book *Principles of Economics* [77], which laid the foundation for neoclassical economics. Relevant to CBA, he also refers to the "Green Book", as a concept of economic progress and social welfare that include access to open spaces and recreational facilities as crucial components of a thriving society. He believed that economic progress should not solely be measured by material wealth but also by improvements in human health, education, and overall quality of life. But it was not before the 1930s when the principle of CBA was newly proposed in the US and formalized for public-sector application. In the United States, the Flood Control Act of 1936 mandated that the benefits of federal infrastructure projects must exceed their costs, as a key principle on how to evaluate ideas in this space [78]. The concept was then spread through U.S. Army Corps of Engineers and U.S. federal agencies. Scholars like Otto Eckstein and Ezra J. Mishan provided theoretical foundations, with Mishan's 1971 textbook becoming a standard reference [79]. CEA emerged slightly later, first in the U.S. Department of Defense, and then also in the healthcare contexts. The key challenge in both contexts is that benefits are difficult to monetize. Economists like Burton Weisbrod in the 1960s [80], and later Weinstein and Stason in the 1970-80s [81], helped define CEA's methodological framework and laid the groundwork for using Quality-Adjusted Life Years (QALYs) in health evaluations. Today, both

CBA and CEA are central to frameworks such as Health Technology Assessment (HTA) and are widely used by organizations like NICE (UK), WHO-CHOICE, and global health funders.

### 2.2.2. Net Present Value (NPV) and Internal Rate of Return (IRR)

In the realm of finance and investment, when evaluating new ideas, two metrics stand out as foundational tools for assessing their feasibility and profitability: Net Present Value (NPV) and Internal Rate of Return (IRR). NPV was formalized by the economist Irving Fisher in his 1907 book *The Rate of Interest* [**82**]. Fisher was an American economist, statistician, inventor and progressive social campaigner, educated and working at Yale University. He introduced the concept of discounted cash flow and the time value of money, forming the basis of modern investment analysis. His 1930 follow-up, *The Theory of Interest*, he expanded on these ideas [**83**]. The IRR, while grounded in similar logic, became prominent in the 1950s through the work of Joel Dean, an American economist and one of the founders of business economics, who worked at several universities in the US. He introduced the IRR to business decision-making in his book *Capital Budgeting* in 1951 [**84**]. Previously, the roots of IRR can be traced to at least two distinct origins. Eugen von Böhm-Bawerk, an economist and Federal Minister of Finance of Austria discussed maximizing net cash flow for each invested currency unit in his book *Positive Theorie des Kapitales* in 1889 [**85**] Then, John Maynard Keynes, an English economist and philosopher whose ideas fundamentally changed the theory and practice of macroeconomics and the economic policies of governments, had discussed similar concepts. He used the term "marginal efficiency of capital" in his *General Theory* in 1936, linking IRR to investment expectations [**86**]. These tools remain core to capital budgeting, used extensively in both private-sector finance and public-sector project appraisal. They also complement methods like CEA and CBA in multi-criteria decision-making.

### 2.2.3. Patent Metrics: Citations and Originality Index

Patent analysis offers another window into the evaluation of novel ideas, particularly in the domains of innovation economics and technology development. While Eugene Garfield focused on scientific literature, his influence permeated the development of patent metrics. Patent examiners at the US Patent and Trademark Office (USPTO) began using citation cards in 1947, which were an early foundation for metrics such as forward citation counts, originality, and generality indexes. Building on Garfield's foundational work in scientific citation indexing [**87**], researchers in the 1980s and 1990s began to apply similar techniques to patents. The aim was to measure the novelty or originality of an invention based on its patent citations, which became widely recognized as indicators of technological impact. This was promoted by researchers such as Adam Jaffe, Manuel Trajtenberg, and Bronwyn Hall. In the paper in 1993, Jaffe et al. demonstrated that the number of times a patent is cited by subsequent patents, i.e. forward citations, correlates with its technological significance and influence [**88**]. In a subsequent 2001 working paper, Hall, Jaffe, and Trajtenberg introduced two additional patent metrics: the *originality index*, which measures how diverse the sources of a patent's knowledge base are; and the *generality index*, which captures how widely a patent is cited across technological fields [**89**]. These metrics, important for evaluation of novel ideas in innovation, are now embedded in patent analytics platforms. They are routinely used by economists, venture capitalists, and technology strategists. It is also worth noting that they are embedded in the databases such as NBER Patent Citations Data File, USPTO, EPO, and OECD REGPAT; they are also used in commercial tools such as Derwent, IFI CLAIMS and Google Patents.

**2.2.4. Citation Metrics, Impact Factor, and h-index**

The assessment of scientific influence and research productivity has evolved significantly over the past century, largely through the introduction of citation-based metrics. These tools, developed independently across several decades, have shaped the way ideas, research impact, and scholarly productivity are evaluated in academia, publishing, and funding landscapes. The foundations of citation analysis were laid by Eugene Garfield, an American linguist and information scientist, often considered the father of scientometrics. In his seminal 1955 paper, *Citation Indexes for Science: A New Dimension in Documentation*, Garfield introduced the idea that scholarly influence could be traced through citations [**87**]. This laid the groundwork for the Science Citation Index (SCI), which was launched in the 1960s and allowed for systematic tracking of idea diffusion through scientific literature. Garfield later established the Institute for Scientific Information (ISI), which played a central role in the development of citation metrics. Garfield, together with Irving H. Sher, later introduced the Journal Impact Factor (JIF), which was formally published in *Science* in 1972 [**90**]. This metric evaluates a journal's average citation frequency within a defined window, initially two years but more recently five years, and has become a widely used - and debated - proxy for journal quality and scientific prestige. Garfield devised the JIF as a tool to primarily assist librarians to evaluate journal quality and influence. Half a century after Garfield's work, Jorge E. Hirsch, a physicist at the University of California, San Diego, introduced the h-index in 2005 [**91**]. This metric aimed to capture both the productivity and the citation impact of an individual researcher's body of work. A scholar has an h-index of *h* if they have *h* publications each cited at least *h* times. The metric gained rapid traction in academic evaluations due to its intuitive appeal and simplicity. These three citation-based metrics - Citation Count, Impact Factor, and H-index - are now embedded in research assessment frameworks, promotion and tenure decisions, grant reviews, and institutional rankings. However, their limitations have also sparked movements such as the San Francisco Declaration on Research Assessment (DORA) and the Leiden Manifesto, advocating for more nuanced and responsible approaches to evaluating scientific contributions [**92**].

**2.2.5. Technology Readiness Levels (TRLs)**

The Technology Readiness Levels (TRLs) framework was pioneered by NASA in the 1970s to evaluate the maturity of technologies intended for space exploration. Originally developed by Stanley Sadin, then Director of NASA's Advanced Projects Office, the system allowed decision-makers to assess how far a particular technology had advanced along its development path [**93**]. The earliest internal use of TRLs occurred between 1974 and 1977. The framework was refined and standardized in subsequent years, culminating in a pivotal 1995 white paper by John C. Mankins titled *Technology Readiness Levels*, which became the definitive articulation of the nine-level TRL scale [**94**]. This structured approach ranges from TRL 1, representing basic research, to TRL 9, indicating fully operational and proven systems. Initially a tool for NASA, TRLs were later adopted by a range of public-sector agencies, including the U.S. Department of Defense, Department of Energy, and the European Space

Agency. The European Commission integrated the framework into its Horizon 2020 and Horizon Europe funding programs. Today, TRLs are used broadly across industries including aerospace, health technology, energy, and manufacturing. The nine levels that were used to evaluate ideas were: TRL 1, in which basic principles were observed; TRL 2, in which technology concept was formulated; TRL 3, with experimental proof of concept; TRL 4, with laboratory validation of components; TRL 5, with validation in relevant environment;  TRL 6, with system/subsystem demonstration; TRL 7, with a prototype demonstration in operational field; TRL 8, with the actual system completed and qualified; TRL 9, with system proven in real-world operational use [**94,95**]. The TRL system has become a global benchmark for technology maturity assessment, guiding investment strategies, research and development funding, and innovation management.

## 2.3. Value-Driven Approaches: Scoring Models and Criteria-Based Frameworks

### 2.3.1. Multi-Criteria Decision Analysis (MCDA)

Multi-Criteria Decision Analysis (MCDA), or Multi-Criteria Decision Making (MCDM), is a formalized family of methods designed to assist in decisions involving multiple, often conflicting, criteria. Intellectual foundations of MCDA trace back to Enlightenment thinkers like Benjamin Franklin's "moral or prudential algebra" in 1772, although it was primarily a method for comparing two alternatives [**96**]. However, the field was shaped into its modern form in the mid-20th century through advances in mathematics, decision theory, operations research, and utility modeling. Research by Harold W. Kuhn and Albert W. Tucker, mathematicians from Princeton University, laid groundwork for MCDA in 1951 [**97**]. The work by Howard Raiffa and Robert Schlaifer at Harvard Business School was instrumental in shaping decision analysis in its current form [**98**]. Bernard Roy pioneered one of the first MCDA methods, ELECTRE (Elimination and Choice Expressing Reality) in France in 1968 [**99**]. Another approach that falls into this family of methods is Thomas Saaty's Analytic Hierarchy Process (AHP), launched in the United States in the 1970s [**74,75**]. Simultaneously, Keeney and Raiffa's 1976 treatise *Decisions with Multiple Objectives* laid the theoretical groundwork for Multi-Attribute Utility Theory (MAUT). This was considered a key progress that highlighted the relevance of multi-attribute utility theory to practical decision analysis, so it continues to underpin many MCDA tools today [**100**]. In 1979, Stanley Zionts' article, *MCDM - If not a Roman Numeral, then What?*, contributed to popularizing the acronym MCDM (for Multi-Criteria Decision Making), and also MCDA [**101**]. In early 1980s, the joint work of Stan Zionts and Jyrki Wallenius on interactive multi-objective linear programming was also influential [**102**]. From 1980s to this date, the field continued to develop with various MCDA methods emerging and being applied to a wider range of complex decision problems, MCDA has evolved to encompass diverse approaches – more than 40 of them are listed on the MCDA's Wikipedia page [**103**]. MCDA is useful in evaluating ideas on how best to address challenges where decisions involve multiple, often conflicting, objectives. It provides a framework for structuring complex problems, assessing alternatives, and evaluating preferences. It evolved over time, building upon earlier ideas and expanding its scope to address increasingly complex decision-making scenarios. Today, it became a broad family of methods used to assess and rank competing ideas, interventions, or alternatives based on multiple conflicting criteria. It was applied in various fields, including public decision-making, resource allocation, project evaluation, public health energy policy, infrastructure planning and technology forecasting. Organizations such as the World Health Organization, NICE (UK), and the European Commission routinely apply MCDA in funding, priority-setting, and evaluation.

### 2.3.2. SWOT Analysis (Strengths, Weaknesses, Opportunities, Threats)

SWOT analysis, perhaps one of the most widely used strategic evaluation tools in business, policy, and project planning, was developed in the 1960s at the Stanford Research Institute (SRI). Its development was led by a business and management consultant Albert S. Humphrey. Originally coined as SOFT (Satisfactory, Opportunity, Fault, Threat), the method was later refined into the now-familiar SWOT (Strengths, Weaknesses, Opportunities, Threats) framework [**104**]. Although no single academic paper formalized the method at the time, Humphrey's approach gained traction through internal corporate planning seminars and practitioner literature. Its aim was to help Fortune 500 companies improve strategic planning by better understanding internal capabilities and external market conditions. Its core innovation was precisely in bridging internal factors – i.e., strengths and weaknesses - with external conditions, which were grouped into opportunities and threats. This enabled organizations to craft strategies that were both self-aware and environmentally responsive. Therefore, SWOT enabled evaluation of longer-term strategic ideas for the future development of companies. A retrospective account by Humphrey, published posthumously in the *SRI Alumni Newsletter* in 2005, recounts how the technique emerged during efforts to improve long-range planning within Fortune 500 companies [**105**]. A parallel lineage of thought could also be found in the influential Harvard Business School text *Business Policy: Text and Cases* by Learned et al. in 1965 [**106**], which laid the groundwork for internal and external analysis in strategic thinking, although it did not yet use the SWOT acronym. Other notable frameworks for strategic idea evaluation include PESTEL and Porter's Five Forces (see later in Chapter 3).

### 2.3.3. Weighted Scoring Models

Weighted scoring models, also known as multi-criteria decision analysis (MCDA) tools, emerged in parallel with many MCDA methods. They emerged during the mid-20th century through the confluence of operations research, decision theory, and systems engineering. These models provided a structured way to evaluate competing alternatives against multiple weighted criteria, especially when trade-offs were necessary and the stakes were high. They were formally introduced by Stanley Zionts in 1979 [**101**], whose mathematical model aimed to aid evaluation process in decision-making by comparing and ranking alternatives based on various criteria. Pivotal progress in applications in decision theory and value-focused thinking was contributed by Howard Raiffa and Ralph L. Keeney from Harvard University's International Institute for Applied Systems Analysis (IIASA) in 1976 [**98**], while the roots of application in operations research reach back to the work by C. West Churchman, Russell Ackoff and E. Leonard Arnoff from the Wharton School at the University of Pennsylvania in 1957 [**107**]. Those three authors are often credited for introducing systems thinking and the idea of assigning weights to criteria in complex evaluations. Stanley Zionts and Barry Boehm helped bring these methods into applied domains, such as engineering, software evaluation, defense procurement and project evaluation in technology settings. Boehm's *Software Engineering Economics* in 1981 explicitly incorporated weighted scoring models into technology decision-making frameworks [**108**]. A typical structure of a weighted scoring model involves: (i) defining decision alternatives; (ii) selecting criteria relevant to evaluation; (iii) assign weights to each criterion, based on importance; (iv) scoring each alternative against each criterion; (v) calculating weighted sum of scores for each alternative; (vi) ranking alternatives based on total weighted score. These kinds of models became popular in project selection, product evaluation, and research and development portfolio analysis [**108,109**].

### 2.3.4. Pugh Matrix (Decision-Matrix Method)

The Pugh Matrix - also known as the decision-matrix method - was introduced by Stuart Pugh, British mechanical engineer and professor of design at the University of Strathclyde in the late 1980s. It was a systematic tool for evaluation of ideas in engineering and product development by comparing alternatives in design concepts, ultimately leading to selecting the best option. It enabled teams to evaluate multiple options against a set of predefined criteria, but relative to a baseline. Pugh formalized this approach in his 1990 book *Total Design: Integrated Methods for Successful Product Engineering* [**110**]. His method emphasizes comparative, and not absolute assessment, using simple symbols (+, –, S) to indicate whether an option performs better, worse, or the same as a reference solution. Pugh's approach helps teams avoid premature convergence on a single idea and fosters objective dialogue in innovation processes. The eventual tally scores guide decisions, but discussion and insight can matter more than raw totals. This makes the matrix particularly useful in engineering, manufacturing, healthcare, and research and development portfolio selection. Avoiding early overcommitment to a single idea and encouraging iterative refinement through team-based analysis are this method's strengths. It has since become a staple of Six Sigma (particularly in the DMAIC framework), Lean product development, and innovation portfolio management.

### 2.3.5. Design Thinking: Feasibility–Desirability–Viability (FDV) Framework

The Feasibility–Desirability–Viability framework for evaluation of ideas in design, commonly known as FDV, was popularized by the global design and innovation consultancy IDEO in the late 1990s and early 2000s. Although no formal academic origin underpins this triad, it emerged organically from IDEO's human-centered design practice and design thinking, described in Chapter 1 (see earlier). It was institutionalized through its teaching at Stanford's d.school, formally Hasso Plattner Institute of Design. The framework assists teams to evaluate innovation ideas through three essential lenses: (i) *feasibility*: can it be built with current technology and skills?; (ii) *desirability*: do people want it?; and (iii) *viability*: is it economically and organizationally sustainable? Thus, the framework emphasizes balancing what is desirable from a user's perspective, what is technologically feasible, and what is economically viable. Tim Brown's 2009 book *Change by Design* [**50**] serves as the primary reference for FDV, integrating insights from earlier systems thinkers like Buckminster Fuller [**111**] and Herbert Simon [**48**], as well as management theorists such as Peter Drucker [**112**], Alexander Osterwalder and Roger Martin [**113**]. While conceptually simple, FDV's strength lies in forcing innovation teams to balance user needs, technical constraints, and financial realities.

## 2.4. Democratic Approaches: Crowd-Based Assessment

### 2.4.1. Wisdom of the Crowd Techniques

The concept of the "wisdom of the crowd" evaluates ideas by non-expert voting. It is rooted in the notion that collective judgments can outperform those of individuals. This approach has deep intellectual roots. It was popularized in its modern form by journalist James Surowiecki in his 2004 book *The Wisdom of Crowds* [**114**]. Surowiecki outlined the conditions under which crowd-based judgments tend to be superior: diversity of opinion, independence of individuals, decentralization of decision-making, and the existence of an aggregation mechanism. The principle was first empirically observed in 1907 by Francis Galton, an English polymath and the early pioneer of behavioural genetics [**115**]. Galton analysed 787 guesses by fairgoers estimating the weight of an ox. The median estimate was remarkably close to the actual weight, suggesting that the aggregate judgment of a diverse group could be quite accurate. Galton's findings were published in *Nature* under the title *Vox Populi* [**115**]. In contemporary contexts, evaluation of various ideas using the "wisdom of the crowd" is operationalized through platforms and processes such as ranked voting, upvoting systems, prediction markets, and participatory idea selection tools. It has found applications in innovation challenges, participatory budgeting, open policymaking, and deliberative democracy. The "wisdom of the crowd" principle led to the development of crowdsourced idea evaluation platforms, including voting-based idea platforms such as IdeaScale, Google Moderator, Dell IdeaStorm and crowdsourced research funding and prioritization, such as OpenIDEO and Foldit. They invite broad participation and aggregate rankings to identify promising ideas [**116**]. Voting mechanisms may include simple upvotes/downvotes, ranked-choice ballots, dotmocracy, or pairwise comparisons—each serving as an aggregation mechanism that aligns with Surowiecki's criteria for collective intelligence [**116**].

### 2.4.2. Prediction Markets

Prediction markets, also known as "information markets" or "idea futures", offer a more formalised mechanism for aggregating collective judgments. These markets allow participants to trade contracts based on the likelihood of future events, effectively translating beliefs into prices that reflect the perceived probability of outcomes. The conceptual foundation of prediction markets was laid by Robin Hanson, an economist and polymath at George Mason University and formerly Caltech/NASA, in the late 1980s. In his 1990 article *Could Gambling Save Science?*, Hanson proposed "idea futures" as a method for forecasting scientific outcomes [**117**]. His later work further refined the mathematical underpinnings of prediction markets, including combinatorial betting and market scoring rules [**118**]. One of the earliest real-world implementations was the Iowa Electronic Markets (IEM). University of Iowa team launched it in 1988, originally focused on political forecasting, aiming to forecast U.S. election outcomes [**119**]. Economists James Wolfers and Eric Zitzewitz contributed empirical validations of prediction markets' accuracy across domains [**120**]. Prediction markets have since been adapted to assess R&D portfolios, evaluate the viability of new technologies, and forecast policy impacts. Companies like Google, HP Labs, Eli Lilly and Microsoft have experimented with internal prediction markets to inform strategic decisions [**121**].

### 2.4.3. The James Lind Alliance (JLA) priority setting partnerships (PSPs)

The James Lind Alliance (JLA) follows a transparent, inclusive, and stepwise process to evaluate proposed research ideas, which are typically framed as "uncertainties". After they are gathered from patients, carers, and clinicians, they undergo a structured evaluation and prioritisation process. In the first step, the Priority Setting Partnership (PSP) performs data collation and theming: all submissions, which may number in the hundreds, or even thousands, are collated and subjected to a qualitative data synthesis. This results in grouping of similar or duplicate questions, removing out-of-scope entries, and merging overlapping questions into broader, clearly phrased indicative questions, also involving thematic analysis and deductive coding. In the second step, checking for 'true' is performed by checking each proposed research idea against the existing evidence to determine whether it is still unanswered and removing those that are. In the third step, all validated research uncertainties are presented in an interim prioritisation survey, which is distributed again to patients, carers and clinicians. They are asked to select the questions they consider most important. Responses are quantitatively analysed, often by calculating frequency of selection, sometimes stratified by stakeholder group to preserve balance between professional and public voices. Typically, the top 25–30 most frequently selected questions proceed to the final prioritisation workshop - a face-to-face or virtual consensus workshop, facilitated using nominal group techniques: mixed small group discussions, bringing together patients, carers and clinicians; then, iterative ranking rounds, where each group discusses and re-orders the questions. In a final plenary session, all groups' rankings are combined, debated, and agreed upon. The outcome is a Top 10 list of jointly agreed research priorities, reflecting a consensus across all stakeholder perspectives. Through this approach, the key Principles that guide evaluation of proposed research ideas are equity of voice for all stakeholders, transparency of the process, grounding in existing evidence, and consensus-building, where deliberative methods are used to arrive at shared priorities [**40,122,123**].

### 2.4.4. The Child Health and Nutrition Research Initiative (CHNRI) Method

The approach to evaluating proposed research ideas in the CHNRI method has been initially described in the CHNRI method's guidelines for implementation in 2008 [**42**]. Further refinements were detailed by Igor Rudan in the fourth paper of the series that updated CHNRI method in 2016, and which explained the method's key conceptual advances [**124**], and later in his book *Measuring ideas: The CHNRI method* published in 2022 and co-edited with Sachiyo Yoshida from the World Health Organization, Kerri Wazny from the University of Edinburgh, and Simon Cousens from the London School of Hygiene and Tropical Medicine [**125**]. Within a chosen field of health research, in the first step, the method would define the existing context and identify a small set of criteria that would recognise one research idea as better than another. These criteria could have been, e.g., likelihood that the proposed research idea would be answerable, that it would lead to effective, deliverable, affordable and/or sustainable health intervention, that it could reduce large portion of the existing disease burden, and that its effects would be equitable. In the next step, dozens of the leading experts in the chosen field of health research would be invited to propose several of their best research ideas. Then, a consolidated list of research ideas is developed after merging ideas that were very similar or removing the duplicate ideas. In the third step, all the experts "score" the ideas by assessing their likelihood of satisfying each criterion - answering simply "yes" or "no". They are also allowed to use an "informed maybe", or to simply leave the scoring field blank, if they didn't have enough knowledge to make this judgement. The preference of "yes"/"no" answer over larger scales with several increasing options prevents regression of the scores to the mean value, but "maybe" is still allowed if it is well-informed, while "blank" option serves to minimise the noise if the scorers are uninformed. As a result, the collective opinion of dozens of leading experts in the field of health research about several hundreds of systematically gathered research ideas are scored through this expert-sourcing, leading to a simple table that shows the "collective optimism" of the experts towards how it would satisfy each of the criteria. All scores are intuitive, as they range between 0-100%, and there are statistical tools that allow computing agreement statistics for each idea, as well as 95% confidence intervals for all scores. This scoring approach offers several advantages over other priority-setting methods. It ensures transparency, as all evaluations follow a clear structure. Also, it minimizes personal biases by relying on a structured survey process rather than subjective discussion. It provides a systematic, reproducible method with well-defined outcomes. Furthermore, it generates intuitive, quantitative scores that can be used for ranking and funding decisions [**124,125**].

**2.4.5. Social Media Engagement Metrics as Proxies for Idea Traction**

With the advent of Web 2.0 and the rise of participatory digital platforms, social media engagement metrics, such as likes, shares, comments, retweets, upvotes, and follower growth, began to serve as informal proxies for gauging the public traction of ideas, content, or innovations [**126**]. While not initially designed for this purpose, such metrics became increasingly influential in decision-making within marketing, science communication, and innovation ecosystems. The period between 2004 and 2009 was marked by early emergence of social platforms, such as Facebook, Twitter and Reddit. The formalization of engagement metrics as indicators of idea diffusion began in the 2010s. Jason Priem, a PhD student at the University of North Carolina at Chapel Hill, and colleagues initiated the altmetrics movement with their *Altmetrics Manifesto* in 2010 [**127**]. They proposed that online interactions, i.e., tweets, blog posts, bookmarks, and other similar expressions on the internet, could complement citation counts in capturing scholarly impact. In parallel, marketing science contributed robust empirical analyses. Katie Delahaye Paine's book *Measure What Matters: Online Tools for Understanding Customers, Social Media, Engagement, and Key Relationships* in 2011 is often cited as one of the key references in this vast new field [**128**]. Paine is a prominent figure in the field of communications and measurement, and her book focuses on using data to understand and improve public relations, social media, and communication strategies. At the Wharton School, professors of marketing Jonah Berger and Katherine Milkman published a study in 2012 entitled *What Makes Online Content Go Viral?* They demonstrated that emotional resonance, novelty, and social currency drive engagement and diffusion, making these signals useful for predicting the reach of new ideas [**129**]. Professor Igor Rudan and his PhD student with a background in music and media industries Iain H. Campbell, from the University of Edinburgh, experimented with a series of videos themed on global health in 2017-2019. They used multiple media outlets to learn more about what themes cause virality among which audience and on which platform [**130,131**]. In 2018, Anatoli Colicev, a scholar in marketing at the Bocconi University and his co-workers developed specific social media metrics, including engagement, to study the impact of social media on brand awareness, purchase intention, and customer satisfaction [**132**].

Engagement metrics are now routinely used in many areas – marketing analytics, research dissemination analytics, technology adoption, innovation trend forecasting, public health messaging and campaign evaluation, and early-stage evaluation of product or policy ideas. However, limitations to their use remain, because "engagement" is a complex construct with various dimensions, such as likes, comments, shares, or time spent on content, that all need to be considered. Also, engagement data can be manipulated by bots or coordinated campaigns, so it may not correlate with long-term impact or quality. Researchers have urged caution, emphasizing the importance of linking engagement to tangible outcomes [**131,132**].

## 2.5. Historically Useful Principles: Scientific and Philosophical Validity Tests

### 2.5.1. Logical Consistency and Deductive Reasoning

The principles of logical consistency and deductive reasoning represent some of the earliest formal tools for evaluating the basic validity of ideas. These concepts were systematically developed by the Greek philosopher Aristotle in the 4th century BCE. Widely regarded as the "father of logic," Aristotle's work laid the foundation for much of Western rational inquiry, influencing disciplines ranging from philosophy and mathematics to modern science. In his treatise *Prior Analytics*, around 350 BC, Aristotle introduced syllogistic logic. This was a formal system in which conclusions are derived logically from two premises. It emphasized that a valid argument, or syllogism, must be logically consistent, for example: *"All humans are mortal. Socrates is a human. Therefore, Socrates is mortal."* Such reasoning exemplifies deductive logic, where the truth of the conclusion necessarily follows from the truth of the premises [**133**]. Aristotle also emphasized logical consistency, notably in *Metaphysics* (Book IV), where he formulated the law of non-contradiction: "It is impossible for the same thing to both belong and not belong to the same thing, in the same respect, at the same time". His methods included *reductio ad impossibile* - testing an argument by assuming the opposite and showing it leads to a contradiction. These ideas became the basis for assessing coherence and truth in any argumentation [**134**]. Aristotle's logical system profoundly influenced the development of Western thought and remained influential for centuries, shaping how people reasoned about ideas and assessed arguments. The principles of logical consistency and deductive reasoning for assessing ideas are among the oldest formal methods in intellectual history. They influenced medieval scholasticism (e.g., Thomas Aquinas), modern logic and set theory (e.g., Gottlob Frege, Bertrand Russell, and Kurt Gödel), and scientific method, where deductive reasoning is used to test hypotheses for internal coherence.

### 2.5.2. Empirical Testability, Replicability and Falsifiability

The criteria of empirical testability, replicability and falsifiability are central to modern scientific method and evaluation of scientific hypotheses. These principles assert that for an idea, hypothesis or theory to be considered scientific, it must be subject to empirical observation and capable of being tested under controlled conditions. Furthermore, the results must be replicable by independent researchers using the same methods. Early advocates of empirical inquiry, most notably Francis Bacon, philosopher and former Lord High Chancellor of Great Britain, emphasized observation and induction in *Novum Organum* in 1620 [**135**]. His work was followed by David Hume, who raised concerns about induction and causation, leading to the need for reproducibility in his book *A Treatise of Human Nature* in 1739 [**136**]; then, Claude Bernard, who emphasized the need for experimental verification in physiology in 1865 [**137**]; and John Stuart Mill, who included the methods of agreement and difference, anticipating modern experimental design, in his book *A System of Logic* in 1843 [**138**]. However, it was Karl Popper, an Austrian philosopher of science, later based in the UK, who gave these criteria their most influential formal articulation. In *The Logic of Scientific Discovery* (originally *Logik der Forschung*, 1934) [**139**], Popper proposed the key principles for evaluation of scientific ideas: empirical testability, i.e., that an idea must be observable and testable; replicability, i.e. that independent tests must yield consistent results; and falsifiability, i.e., that it must be possible to disprove the idea;. These criteria became central to scientific method, hypothesis testing, clinical trials, reproducibility crisis discussions, and philosophy of science. Popper considered falsifiability the defining criterion of science. He argued that science progresses not through verification, but through conjecture and refutation. Therefore, scientific theories must be structured in a way that they can, in principle, be proven false, and bold hypotheses must expose themselves to possible disproof. For a theory to be scientific, it must predict what should not happen and then be testable in a way that observations can potentially contradict it. Popper also highlighted the importance of replicability as a safeguard against bias, error, or coincidence, demanding that independent researchers should be able to reproduce results under the same conditions. Still, he underscored that confirming a theory through repeated observation is inherently weaker than designing tests to falsify it [**139**]. The criteria of testability, replicability and falsifiability offered a logical solution to the philosophical problem of demarcation: how to distinguish between science and non-science. A theory is scientific only if it makes specific, risky predictions that exclude certain outcomes. If no conceivable observation could contradict the theory, then it is unfalsifiable, which makes it unscientific. Popper cited Einstein's theory of relativity as a model of falsifiability, since it made precise predictions that could be tested and refuted. In contrast, he viewed Freud's psychoanalysis and Marxist historicism as pseudoscientific because they could accommodate any outcome and therefore resisted falsification.

### 2.5.3. Paradigm Shifts

In contrast to Popper's focus on individual hypotheses and their testability, Thomas S. Kuhn, an American historian and philosopher of science who worked at several US universities - Harvard, Berkeley, Princeton and MIT - offered a sociological and historical perspective on how scientific ideas evolve. In his groundbreaking book *The Structure of Scientific Revolutions* in 1962, Kuhn introduced the concept of a paradigm shift. This is a radical, collective reorientation in the basic assumptions, methods, and questions within a scientific community [**140**]. According to Kuhn, science proceeds through long periods of traditional research, during which researchers solve puzzles within an established paradigm. Over time, however, anomalies accumulate and generate an increased amount of data that the prevailing theory cannot explain. Eventually, a crisis emerges, culminating in a scientific revolution and the adoption of a new paradigm. This new paradigm is often incommensurable with the old; in some cases. His concept of "incommensurability", suggested that different paradigms may be fundamentally incompatible and cannot even be directly compared. Examples cited by Kuhn include the transition from Ptolemaic to Copernican astronomy, from Newtonian physics to Einstein's relativity, from classical chemistry to atomic theory, and from classical genetics to molecular biology. Therefore, Kuhn argued that scientific progress isn't a linear accumulation of facts. Instead, it occurs through revolutionary shifts between different paradigms. His work challenged the traditional view of science as a purely objective and cumulative process, emphasising the role of social and historical context in shaping scientific knowledge, professional consensus, and sociocultural factors. Essentially, against Popper's argument that ideas should only be judged by falsifiability or logical consistency, he noticed that they are also assessed within the context of dominant paradigms, which define what counts as evidence, what questions are worth asking, and what methods are valid. Thus, a paradigm shift changes the entire assessment framework, where once rejected concepts can become foundational, and vice versa.

## 3. Idea Prioritisation Methods

### 3.1. Rational Decisions Through Structure: Structured Decision-Making Frameworks

#### 3.1.1. Paired Comparison Methods

Paired comparison methods enable decision-makers to prioritise ideas by comparing them two at a time. They have a long history in psychology, decision theory, and preference elicitation. Louis L. Thurstone, the American psychologist from the University of Chicago and pioneer of psychometrics, is credited for introducing them formally in the 1920s as part of psychometric scaling. In his 1927 paper *A Law of Comparative Judgment* [**141**], he demonstrated that this cognitively simple, yet powerful approach can be useful in investigating a wide range of psychological attributes, such as "seriousness of crime". His work allowed for the derivation of quantitative preference scales from simple ordinal choices. By presenting respondents with pairs of alternatives and asking them to choose the preferred one, paired comparison methods create a matrix of preferences from which overall rankings can be statistically inferred. Thurstone introduced the statistical model behind paired comparisons and derived interval-scale values for preferences based on how often one item is chosen over another. These models, such as Thurstone's and later the Bradley-Terry [**142**] and Elo models [**143**], are now foundational in decision science and preference elicitation. Paired comparison remains particularly valuable when stakeholders are uncomfortable assigning numerical weights, but can reliably express ordinal preferences. Modern frameworks like AHP and certain Delphi adaptations often integrate paired comparisons as a core or optional feature. Paired comparison methods are used to prioritise health research (e.g., James Lind Alliance, Delphi hybrid methods), educational learning needs, ideas in marketing, product development, military and intelligence analysis, and for crowdsourced prioritization, such as voting systems.

#### 3.1.2. Multi-Voting and Dot Voting

Multi-voting, also known as "Dot Voting", "Dotmocracy", or "Sticker Voting", is a simple and highly adaptable method used to identify priorities in group settings. It likely originated informally in the 1950s–1960s in industrial and educational group facilitation and grew out of group facilitation practices. Then, it was popularized in the 1970s–1980s by practitioners of Total Quality Management (TQM), Six Sigma, facilitation training programs, and later by design thinking and agile teams. Because of this, it has no single credited inventor. Later, it became a standard tool in design thinking, quality management, and Agile project management. In multi-voting exercises, participants are given a fixed number of "votes". These can be either dots, stickers, checkmarks, or online tokens. Then, the participants allocate their votes among competing ideas that are visually displayed to them - usually on flipcharts, whiteboards, or similar surfaces. The votes are then tallied to generate a prioritized list based on collective preference. This method's appeal lies in its simplicity, speed, and inclusiveness. It is now widely featured in facilitation manuals, such as Bens' *Facilitating with Ease* [**144**], continuous improvement guides like Brassard's and Ritter's *The Memory Jogger II* [**145**], and IDEO's and Stanford d.school's design toolkits that were mentioned earlier.

### 3.1.3. Nominal Group Technique with Voting (NGT)

The Nominal Group Technique (NGT), a structured, face-to-face method for idea generation and prioritisation using voting was introduced in the late 1960s and early 1970s by Andre L. Delbecq, Andrew H. Van de Ven, and David H. Gustafson. They were American management scholars and worked at the Universities of Wisconsin and Minnesota the time. NGT was designed as a structured method for group brainstorming and idea prioritization. It was designed to ensure inclusive participation and minimize the dominance of strong, outspoken personalities in face-to-face settings, which was a common weakness of many other methods in use at the time. Its approach was to combine individual idea generation with structured group discussion and prioritisation by anonymous voting. Their approach facilitates consensus-building while preserving individual creativity. It was first explained in their book *A Group Process Model for Problem Identification and Program Planning* in 1971 [**35**], followed by another book that compared it to Delphi processes in 1975 [**36**]. Those two books provided a comprehensive guide to the technique, including its application for consensus-building and action planning. NGT outlines a five-step process. In silent idea generation, individuals silently generate ideas in writing; then, in round-robin sharing, each member shares one idea at a time, recorded without debate; in the clarification stage, the group discusses each idea for understanding, but not evaluation; finally, the anonymous voting or ranking takes place, in which participants independently and anonymously vote or rank ideas, often using a point allocation method. The results are then aggregated In the way that votes are tallied to produce a prioritised list. This method preserves individual creativity, encourages equal input, and produces quantifiable prioritisation. [**35,36**]. NGT has been widely used in health research priority-setting, clinical guideline development, educational planning and curriculum design, community planning, strategic planning in government and nonprofit sectors, including the James Lind Alliance and NIHR UK.

### 3.1.4. Delphi Process with Ranking Rounds

The elements of generation and evaluation of ideas within the Delphi process have already been described under previous subheadings [**34,36**]. In this subchapter, we focus on its approach to prioritisation of ideas. In the Delphi method, once expert ratings are collected for each idea, facilitators compute summary statistics, such as the mean, median, interquartile range, or degree of agreement, to quantify how each item was evaluated. This aggregated information is then shared back with the participants in anonymized form, allowing them to see how their views align or diverge from those of their peers. In the subsequent round - typically the third and beyond if needed - participants are invited to review their previous ratings in light of this group summary and may revise their scores. This controlled feedback loop serves to highlight outlier opinions, surface emerging consensus, and encourage convergence toward a collective judgment. When stability in responses across rounds indicates that sufficient consensus has been achieved, the final results are compiled into a ranked list of ideas. This final prioritisation can be based on median or mean scores, or through weighted scoring when some criteria are deemed more important. In more complex cases, composite scoring systems may be used to combine multiple evaluation criteria into a unified index. Some Delphi exercises conclude at this stage, while others proceed to classify the ranked ideas into categorical tiers, such as identifying the top five priorities, or assigning levels of urgency (e.g., high, medium, low), or implementation horizons (short-, medium-, or long-term). To ensure rigor and consistency, many Delphi protocols define consensus thresholds in advance. They may require at least 70% of participants to rate an item 7 or higher on a 1–9 scale or expect an interquartile range (IQR) of no more than IQR=2 to indicate sufficient agreement. These predefined rules help determine which ideas are ultimately retained, discarded, or considered most important. Several methodological features enable this structured ranking and prioritisation process. Anonymity protects participants from dominance bias; iteration allows progressive refinement of views; controlled feedback fosters peer learning; quantitative scoring supports systematic evaluation; and statistical aggregation enables transparent consensus measurement. In cases where the goal is to narrow down a shortlist of priorities, e.g., in funding or policy, a final round may include a forced ranking exercise or a points-allocation system in which participants distribute a fixed number of points across the most promising ideas.

### 3.1.5. Analytic Hierarchy Process (AHP)

The elements of generation and evaluation of ideas within the Analytic Hierarchy Process (AHP), developed by Thomas L. Saaty [**74,75**], the American mathematician and operations researcher from the University of Pittsburgh, have already been described under previous subheadings. In this subchapter, we reflect on its approach to prioritisation of ideas. Saaty sought a rigorous, yet intuitive framework for making structured judgments in the face of competing alternatives, and for selecting rational choices in the face of multiple, conflicting objectives. He broke complex decisions down into a hierarchy of goals, criteria, subcriteria, and alternatives. Once the users of AHP evaluated the relative importance of elements through pairwise comparisons, AHP generates quantitative weights via eigenvalue calculations. It computes a "priority vector" using eigenvalue methods. A consistency ratio is also calculated to check the reliability of these judgments. AHP has since been applied extensively in fields ranging from global health research and public policy to corporate strategy and engineering design.

### 3.2. Addressing Health and Development: Priority-Setting Frameworks in Health Sciences

### 3.2.1. RAND/UCLA Appropriateness Method (RAM)

The RAND/UCLA Appropriateness Method (RAM) was developed in the mid-1980s by researchers at RAND Corporation in collaboration with clinicians at the University of California, Los Angeles (UCLA) to assess the appropriateness of medical interventions and surgical procedures. It was partly motivated by the concerns about overuse, underuse, and variability of the delivered interventions in clinical practice. RAM combines a systematic review of the literature with expert panel judgment. Experts rate medical procedures using a scale of 1 to 9, considering both the evidence base and clinical scenarios. Ratings are analysed for consensus and disagreement, helping to identify and prioritise procedures that are appropriate, inappropriate, or uncertain. The key reference, *RAND/UCLA Appropriateness Method User's Manual*, was published in 2001 following years of experience [**146**]. It formally codified the method and provided step-by-step guidance for its implementation. Since its inception, RAM has been widely applied in healthcare quality improvement, technology assessment, and coverage decisions. It represents a hybrid between empirical data use and expert deliberation.

### 3.2.2. Essential National Health Research (ENHR) Framework

The Essential National Health Research (ENHR) framework was introduced in 1990 by the Commission on Health Research for Development, which later evolved into the Council on Health Research for Development (COHRED). The framework was articulated in the landmark report *Health Research: Essential Link to Equity in Development*, published in 1990 on behalf of the Commission [**147**]. ENHR sought to reorient national research systems in LMICs toward equity-driven, needs-based, and country-led priorities. It emphasized integrating research with health system goals, involving diverse stakeholders, and supporting national capacity for implementation. ENHR became a foundational model for aligning research with development and equity goals and promoting research as a public good, rather than a purely academic enterprise. Although it was received by great enthusiasm by the governments in low- and middle-income countries due to those welcome aims, the uptake and practical application of this framework has been relatively modest to date. In the paper by Sachiyo Yoshida from the World Health Organisation, based on her analysis of the usage of different methodological approach studying PubMed database, ENHR was used in only 0.6% of all the papers that attempted to set health research priorities between 2001 and 2014 [**148**]. This is likely because, although ENHR does well to point to the general, broad research aims and goals for national health systems in terms of country leadership, increased focus on meeting the needs, and improved equity in the population, it is not prescriptive and decisive in picking the priorities as many other methods are. It doesn't end with clear lists and ranks of ideas, but rather with broad and general recommendations. Although these are most often very useful and point the policy makers to the right direction, the output lacking in specificity can later be subject to differences in interpretation among the stakeholders

### 3.2.3. GRADE Methodology and Evidence to Decision (EtD) Frameworks

The Grading of Recommendations Assessment, Development and Evaluation methodology (GRADE) was initiated by the GRADE Working Group in 2000 as a global collaboration to improve the clarity, transparency, and reliability of clinical recommendations [**149**]. This group is a collaborative of over 500 scientists, clinicians, methodologists, and other experts dedicated to a transparent and systematic approach to assessing evidence and developing recommendations. The EtD frameworks specifically are used to make well-informed healthcare choices by systematically evaluating evidence and making transparent decisions. It has since become the dominant approach used by guideline developers, including WHO, NICE, and Cochrane. GRADE distinguishes the certainty of evidence (high, moderate, low, very low) from the strength of recommendations, incorporating factors such as benefits, harms, values, preferences, and resource use. Building on this foundation, the Evidence to Decision (EtD) frameworks were developed during the DECIDE project (2011–2015) [**150**]. These frameworks make explicit how evidence is translated into recommendations by offering structured domains such as equity, feasibility, acceptability, and cost-effectiveness. Over subsequent years, iterative methods that include literature review, stakeholder feedback, and user testing were used to produce EtD frameworks tailored to different decision-making contexts: clinical recommendations, coverage decisions, and health system/public health decisions

### 3.2.4. James Lind Alliance Priority Setting Partnerships (JLA PSPs)

The elements of generation and evaluation of ideas in the JLA PSPs process have already been described under previous subheadings [**40,122**]. Here, we focus on its approach to prioritisation of ideas. Once the uncertainties are validated, they are then organized into longlists and subsequently shortlisted through stakeholder surveys where participants score questions based on their perceived importance. The final stage of prioritisation is carried out in face-to-face consensus workshops using structured, voting-based techniques - typically adaptations of the Nominal Group Technique (see earlier text), where diverse stakeholders deliberate and rank the shortlisted questions. This inclusive, democratic process culminates in a final "Top 10" list of agreed research priorities, which serves as a guidance tool for funders, policymakers, and research organisations. These Top 10 lists are widely recognised for their legitimacy and relevance, having informed funding calls and guideline development in multiple clinical areas. More than a hundred PSPs have been completed across various diseases and care settings, establishing the JLA as a globally influential model for equitable and evidence-based research prioritisation. Its impact has extended internationally, shaping practices adopted by institutions such as the World Health Organization, the Cochrane Collaboration, NIHR UK, MRC UK, and others. Key documentation of the JLA's methods includes the evolving JLA Guidebook [**122**] and a landmark article that provides a conceptual and empirical foundation for reducing waste in research through inclusive priority setting [**151**]. The legacy of the JLA lies in its pioneering approach to shared decision-making in research agenda setting, creating a replicable model for integrating lived experience into the production of evidence and ensuring that scientific inquiry is more responsive, democratic, and impactful.

### 3.2.5. Combined Approach Matrix (CAM)

The Combined Approach Matrix (CAM) was introduced by the Global Forum for Health Research in 2004 as a practical tool for health research priority setting, particularly in low- and middle-income countries (LMICs). Developed under the leadership of Stephen Matlin, an international expert in global health innovation, through technical leadership by Abdul Ghaffar and contribution from Andres de Francisco, both global health and development experts [152], CAM aimed to address the so-called "10/90 gap". This was the observation that less than 10% of global health research resources targeted conditions affecting 90% of the world's population [153]. The CAM integrates multiple dimensions in prioritising health research, aiming to improve the process in which the scientists discuss and decide on funding priorities based on their own views and knowledge. CAM was piloted in countries such as Mexico, Tanzania, and Burkina Faso, where it helped to shift attention to locally owned research agendas. It was useful for systematic classification, organization, and presentation of the large body of information needed at different stages of priority setting process. As a result, the decisions made by the committees were based on information, rather than personal knowledge and judgment. CAM incorporates "economic" dimension along one axis, and "institutional" along the other, thus covering the determinants of health at the population level. Components of the economic dimension are "disease burden," its "determinants," "present level of knowledge," "cost and effectiveness," and "resource flows." Components of the institutional dimension are "the individual, household and community," "health ministry and other health institutions," "sectors other than health," and "macro-economic policies." CAM can be applied at the level of disease, risk factor, group or condition, and also at local, national, or international level [152-154]. In terms of uptake, the paper by Sachiyo Yoshida from the WHO showed that CAM was used in only 1.8% of all the papers that attempted to set health research priorities between 2001 and 2014 [148]. The likely reason is that it did not offer an algorithm or system for ranking or discriminating the competing investment options. Therefore, in the absence of reliable information, which is common in the countries that are in need of tools like CAM, most of the decisions were still based on discussions and agreements within the panels of experts. However, CAM laid methodological foundations for further work of the Global Forum for Health Research on addressing health research priority setting. The later framework, the CHNRI method, which also arose from the Global Forum for Health Research in 2006-08, built on CAM and gradually received global uptake and widespread use in global health research prioritisation.

### 3.2.6. Child Health and Nutrition Research Initiative (CHNRI) method

Building on CAM and several other previous attempts to prioritise health research ideas, the CHNRI method proposed a more structured approach. It led to more than 200 published exercises based on the CHNRI method to date, with some beginning to expand to areas beyond health research. As result of the scoring process described earlier, each proposed research idea receives an intermediate score for every criterion that is used for its evaluation. Each score assesses how well a research idea satisfies a specific criterion - e.g., answerability, effectiveness, deliverability, equity, and others. The intermediate scores are computed by averaging all non-"blank" responses (i.e., "1", "0", or "0.5" points), thus ensuring that missing responses do not distort the calculations. All intermediate scores thus become percentages, providing a standardized measure of collective opinion of many leading experts in the field. At this stage, the management team may decide to place weights or apply thresholds on selected criteria, based on the input from many stakeholders. The overall Research Priority Score (RPS) can then be calculated as the weighted mean of all intermediate scores, after the ideas that didn't meet the thresholds are excluded. The final RPS can then be used for ranking, and the uncertainty of the final RPS for each research idea can be statistically using bootstrapping-generated confidence intervals. This approach to prioritisation offers several advantages. It ensures transparency, minimizes biases and eliminates undue influences among participants. It provides a systematic, reproducible approach with simple, well-defined and intuitive outcomes that can be used for funding decisions [**125,155**]. It also assesses the level of agreement between the scorers, thus exposing the areas of greatest controversy and providing an opportunity for targeted discussions on priorities after the completion of the process. The level of agreement can be assessed by a simple measure of Average Expert Agreement (AEA), but also Kappa statistics, or more recently, an improved version of the AEA score based on information theory, defined as the exponential of the negative entropy. Entropy is a widely used information criterion to quantify uncertainty; in this case, higher entropy implies greater uncertainty and less agreement. The CHNRI method can also assess the internal structure of the scorers through studying the diversity of their responses and identifying clustering. Various methods of hierarchical clustering can be used to this end. One of the most recent advances of the CHNRI method was its adoption of artificial intelligence and Large Language Models (LLMs). In one of the recent exercises, the output of the LLM, based on its training on human collective knowledge, was compared with the results obtained through human collective opinion of the leading experts [**155**]. This fascinating new prospect may assist us to better understand the differences between human and AI-based prioritisation, representing an early example where humans and AI can supplement each other towards a desirable outcome. The CHNRI method also includes two approaches to linking priorities either with specific decisions using a new funding portfolio, or with optimisation of an existing funding portfolio. To ensure that research investments remain cost-effective and aligned with current needs, the methodology allows for periodic re-evaluation and refinement [**125,155**].

### 3.3. Prioritising in Private Sector: Portfolio and Pipeline Management Tools

### 3.3.1. Real Options Analysis (ROA)

Real Options Analysis (ROA) is a method for prioritising innovation investments under uncertainty. Just as a financial option has value, so too does the flexibility to delay, expand, scale, or abandon a real-world project or idea under uncertainty. A tool that takes this into account would assist in prioritising strategic ideas based not only on expected value, but also on the value of waiting or adapting decisions over time. It would, therefore, capture option value in decisions where uncertainty and learning are significant, especially in strategic planning, innovation, research and development, pharmaceuticals, and rapid technology development. The approach extends financial options theory to real-world assets. It emphasises the value of flexibility in decision-making. Essentially, it offers a valuation framework that quantifies the strategic value of flexibility in uncertain environments. The concept was introduced in 1977 by Stewart C. Myers, Professor of Financial Economics at the MIT Sloan School of Management. He argued that traditional net present value (NPV) methods undervalue investments by ignoring managerial flexibility [**156**]. He defined "real options" as "opportunities to purchase real assets on possibly favourable terms". Myers' work focused on the idea that firms can be viewed as having both "real assets" and "real options", where real options represent opportunities to acquire assets on potentially advantageous terms. This approach allowed for the valuation of investments with embedded flexibility, like the ability to expand, contract, or abandon a project based on evolving information, which are not adequately captured by traditional discounted cash flow (DCF) methods. Thus, it became a tool that allowed prioritisation of strategic decisions. The approach was further formalised by Avinash K. Dixit and Robert S. Pindyck, professors of economics at Princeton and MIT universities, in their 1994 book *Investment under Uncertainty*. Their book provided a rigorous economic foundation for the theory [**157**]. Lenos Trigeorgis, professor of strategy and finance at Durham University Business School, then expanded its practical applications, especially for strategic decision-making in research and development, energy, and public policy [**158**].

### 3.3.2. Research and Development (R&D) Portfolio Matrices: BCG and GE-McKinsey

R&D portfolio matrices, such as the Boston Consulting Group (BCG) Matrix and the GE–McKinsey Matrix, were developed in the 1970s as strategic tools to prioritise project or product ideas, research and development ideas, or business units. Their prioritisation was based on key dimensions like "market attractiveness" and "internal capabilities". These matrices were then applied in strategic management and later adapted to research portfolio decision-making in the sectors like pharmaceuticals, technology, and global health, which are heavily dependent on R&D. Portfolio matrices guided strategic investment decisions and became popular in the 1970s. They have since been adapted to manage research, development, and innovation portfolios. The Boston Consulting Group (BCG) Matrix, also known as the growth-share matrix, was introduced in 1970 by Bruce Henderson [**159**]. Its purpose is to help companies allocate resources across a portfolio of businesses or products based on market growth rate and relative market share. It classifies business units or projects into four quadrants — "Stars", "Cash Cows", "Question Marks", and "Dogs", based on two axes: market growth rate (external opportunity) and relative market share (internal strength). It helps decision-makers allocate resources effectively by identifying which units to invest in, maintain, or divest. Furthermore, the GE–McKinsey Matrix was developed in the early 1970s through a collaboration between General Electric and McKinsey & Company [**160**]. Its purpose was to evaluate business units based on industry attractiveness and competitive strength, providing a more detailed approach than the BCG matrix. It introduced greater complexity with nine boxes instead of four, evaluating business units based on industry attractiveness and business strength. By the 1980s, both matrices were adapted for use in research-heavy fields such as pharmaceutical development, technology planning, and global health research funding. They remain influential in identifying which R&D initiatives to scale, monitor, or close. Comparatively, BCG matrix is a 2x2 grid with axes of market growth vs. market share, while GE-McKinsey Matrix is 3x3 grid with axes of industry attractiveness vs. business strength [**161**].

### 3.3.3. Product Roadmapping and Prioritization Grids

Two influential tools for prioritising innovation in dynamic environments are product roadmaps and prioritisation grids. They evolved over time through contributions from product development, technology management, and agile software communities. The concepts were formalized and popularized between the 1970s and early 2000s through academic, industry, and consulting literature. Though developed independently, both help organisations to align innovation initiatives with their strategic goals. Product roadmapping evolved in technology management circles during the 1970s and 1980s. It was academically formalised by Robert Phaal, Clare Farrukh, and David Probert at the University of Cambridge's Institute for Manufacturing [**162**]. Its purpose is to align product and technology development with strategic goals across time horizons. It is a visual timeline that links market drivers, product goals, and enabling technologies across time horizons. It helps coordinate innovation efforts and communicate priorities to diverse stakeholders. Prioritization grids visually compare ideas along two or more criteria, such as impact vs. feasibility or effort vs. value. Their purpose is to compare and rank ideas based on two or more criteria (e.g., Impact vs. Feasibility). They are rooted in quality management, design thinking, and agile product development. Notable contributors to these tools include Noriaki Kano, who introduced the Kano Model to categorise features based on customer satisfaction in 1984 [**163**]. The Kano model is a set of guidelines and techniques used to categorize and prioritize customer needs, guide product development and improve customer satisfaction. Then, IDEO and Stanford d.school popularised effort-impact matrices in innovation workshops. The Lean Startup movement and Agile sprint planning, which integrated prioritisation grids into software product backlogs. Both tools are now widely used in global health strategy, public sector foresight, and corporate innovation planning.

### 3.3.4. Stage-Gate Model

The Stage-Gate model, also known as the Phase-Gate model, was developed in the 1980s by Robert G. Cooper based on empirical studies of high-performing innovation projects at companies such as DuPont, Exxon, and United Technologies. The model was formally introduced in Cooper's seminal 1990 publication, *Stage-Gate Systems: A New Tool for Managing New Products*, in *Business Horizons* [**164**]. The model structures the innovation process into a series of discrete stages, each focused on specific activities such as idea generation, concept testing, development, and commercialization. These stages are separated by decision "gates", where cross-functional teams assess whether the project should advance, be modified, or terminated. Each gate evaluates the project against pre-defined criteria such as strategic alignment, market potential, and technical feasibility. The Stage-Gate model remains a cornerstone of structured innovation, especially in sectors such as consumer goods, pharmaceuticals, and industrial research and development.

### 3.4. Modern Approaches in Digital Age: AI-Driven Prioritization Tools

### 3.4.1. Knowledge Graphs and Semantic Similarity Clustering

Knowledge graphs and semantic similarity clustering have emerged as powerful computational tools for prioritising research ideas by systematically mapping relationships among concepts and identifying high-impact clusters for further exploration. Rooted in early work on semantic networks from the 1960s, such as the models of Allan M. Collins and M. Ross Quillian [**165**], American cognitive scientists from Northwestern University and BBN Technologies. These approaches have matured significantly over the past two decades through advancements in knowledge representation, ontology development, and natural language processing (NLP). A knowledge graph is a structured representation of entities (like diseases, genes, or interventions) and the semantic relationships that connect them. It enables researchers to visualise and traverse complex idea landscapes, helping identify conceptual clusters, detect research gaps, and inform strategic funding. The concept gained widespread attention in 2012 when Google formally introduced its "Knowledge Graph" to improve semantic search [**166**], but the biomedical domain had already laid foundational work with the Unified Medical Language System (UMLS) developed by the US National Library of Medicine in the 1990s [167]. This ontology-driven framework enabled early applications of semantic clustering in health research. Semantic similarity clustering, a complementary approach, uses algorithms to group ideas or terms based on ontological proximity or vector-space similarity, often using tools like MeSH, UMLS, and embedding models derived from deep learning. This technique has been used for deduplicating proposed research questions, thematic mapping of large literature corpora, and identifying emergent clusters that may merit prioritisation. Since 2015, researchers like Percha and Altman have applied these tools to structure biomedical knowledge extracted from unstructured text [168], while Himmelstein et al. demonstrated how graph traversal techniques can prioritise drug repurposing candidates based on the connectivity of evidence [169]. Large-scale initiatives, such as IBM Watson Health [170], Semantic Scholar from Allen Institute for AI [171], Elsevier's research mapping systems [172], Microsoft Academic Graph [173], and the NIH's NCATS Biomedical Data Translator Consortium [174], have operationalised these tools to support automated hypothesis generation, research portfolio analysis, and strategic foresight. From around 2018, these methods began being explicitly applied to idea prioritisation, especially in biomedical sciences where the volume and complexity of ideas are immense. Unlike traditional deliberative or expert-driven methods, knowledge graph-based prioritisation is data-driven, scalable, and less prone to individual bias. It enables automated and reproducible discovery of promising research areas by identifying semantically rich clusters of ideas that align with knowledge gaps or interdisciplinary convergence. Key publications defining this field include those by Ehrlinger and Wöß in 2016 on formal definitions of knowledge graphs [175], Wang and colleagues in 2017 on graph embeddings to compute semantic similarity between concepts and enable clustering [176], Percha and Altman in 2015 on the use of text mining and semantic networks for biomedical idea mapping and hypothesis generation [177], and Himmelstein et al. in 2017 on knowledge integration for drug prioritisation and translational medicine [169]. Together, these works have laid the groundwork for a new class of AI-enabled methods for research prioritisation that are now integral to modern scientific discovery and funding intelligence systems [178].

### 3.4.2. Reinforcement Learning–Based Portfolio Optimisation

Reinforcement learning-based portfolio optimization is a relatively recent advancement at the intersection of artificial intelligence and decision science. It applies reinforcement learning (RL) techniques to dynamically manage and prioritise investment portfolios, including research and development, financial, and diverse idea portfolios within uncertain, dynamic and rapidly evolving environments. In this new approach to prioritisation, Reinforcement Agent learns to take actions in an environment to maximize cumulative reward over time, optimising portfolio to allocate resources (e.g., capital, research funding, idea investment) across options to maximise expected value. Learning (RL) has increasingly been applied to the optimisation of research, innovation, and investment portfolios. While the concept has no single originator, it builds on the convergence of foundational work in reinforcement learning theory, portfolio management, and deep learning. The theoretical underpinnings were established by Richard S. Sutton and Andrew G. Barto, whose 1998 textbook *Reinforcement Learning: An Introduction* remains the standard reference [179]. The first applications of RL to portfolio optimisation were pioneered by Moody and Saffell in 2001 [180], who used policy-gradient methods for financial trading. By the 2010s, researchers such as Jiang et al. applied deep reinforcement learning (Deep RL) to dynamic asset allocation problems, extending the method's relevance to complex, non-linear decision spaces, with a key reference published in 2017 [181]. OpenAI and Google's DeepMind then extended RL to large-scale planning in science and complex decision environments in gaming with examples such as AlphaZero [182]. RL introduced adaptive learning agents that allocate resources or prioritise ideas dynamically based on observed feedback and evolving environments. They enabled self-improving, data-driven prioritisation under uncertainty, making it especially valuable for innovation portfolios where outcomes are difficult to predict and decisions must be staged or revisited. They have emerging use in R&D portfolio management, drug development pipelines, global health prioritisation, and AI-led scientific discovery.

### 3.4.3. Automated Priority Setting via Large Language Models

The application of Large Language Models (LLMs) such as GPT, Claude, Gemini, and DeepSeek to automated research prioritisation is a frontier development that began to crystallise between 2023 and 2024. These models, built on billions of parameters and trained on vast amount of textual sources, are now capable of reasoning across criteria and ranking ideas at scale, which was previously the domain of expert panels and consensus methods. The first peer-reviewed demonstration of LLMs applied to research idea scoring, and its direct comparison to human expert group's scoring that used the CHNRI method, was published in 2024 by the global health experts Peige Song from Zhejiang University, Igor Rudan from the University of Edinburgh, and their colleagues from the International Society of Global Health (ISoGH) [155]. They compared the top research priorities that the ISoGH's global health experts generated using the CHNRI method to those generated by ChatGPT for the challenge of pandemic preparedness. It was an example how human collective opinion can be compared to human collective knowledge that was used for AI LLM's training, with the outcomes indicating the similarities and differences between humans and AI. This approach opened the door for AI-led research foresight, but also rapid priority setting in funders and public health in emergency situations. The approach is highly scalable and cost-effective, allowing the rapid evaluation of a very large number of ideas, particularly useful in time-sensitive or resource-constrained settings [155,183]. It has potential to complement and even augment traditional frameworks like CHNRI, Delphi panels, or multi-criteria decision analysis (MCDA). It is expected that other early adoptors may be Chan Zuckerberg Initiative with their use of artificial intelligence (AI) for scientific discovery [184], the work at the MIT and Stanford in hypothesis generation [185], and Semantic Scholar integrations at the Allen Institute for AI [186].

### 3.5 Engaging Society in Deliberation: Participatory and Democratic Prioritization

### 3.5.1. Cross-Cutting Philosophical and Meta-Theoretical Approaches

Cross-cutting philosophical and meta-theoretical approaches for the prioritisation of ideas are not attributed to a single inventor or publication. They represent a meta-level evolution of thought across epistemology, philosophy of science, and decision theory. These approaches aim to critically assess, compare, and synthesize various idea-evaluation frameworks by exploring their underlying assumptions, value systems, and conceptual boundaries. Therefore, beyond procedural methods described in the rest of this paper, a rich body of philosophical and meta-theoretical work has shaped how we can think today about the prioritisation of ideas, critically examining the epistemological, ethical, and institutional foundations of prioritisation frameworks. These approaches offer insight into how knowledge, values, and power interact in decision-making, encompassing diverse intellectual traditions. They aim to develop meta-criteria, such as robustness, fairness, scalability, and inclusiveness, for choosing among evaluation frameworks. They can examine the limits of quantification, value-ladenness of science, and disciplinary worldviews in prioritisation. They also attempt to reconcile qualitative and quantitative paradigms in evidence and decision-making. The representatives of this line of thought are philosophers of science, such as Thomas Kuhn, Imre Lakatos, Helen Longino, Paul Feyerabend, Nancy Cartwright, and science policy scholars such as Helga Nowotny, Sheila Jasanoff, Silvio O. Funtowicz, and Jerome Ravetz. Their foundational ideas were presented in the three decades between 1960s and 1990s, and applied to idea prioritisation from 1990s onward, especially in post-normal science, research programs, and meta-framework evaluations. Thomas Kuhn, who was already mentioned in Chapter 2, introduced the notion of paradigm shifts and the incommensurability of scientific worldviews, pointing to the issue of paradigm-dependence in how ideas are evaluated and prioritised [**140**]. Imre Lakatos and Paul Feyerabend challenged the idea of singular rational methods, advocating for methodological pluralism. In the 1970s, Imre Lakatos developed the idea of research programmes as evolving systems of ideas, challenging static hypothesis testing [**187**]. Paul Feyerabend advocated for epistemological anarchism, critiquing rigid methodological prioritisation [**188**]. The Biopsychosocial model, a cross-cutting philosophical and meta-theoretical approach for prioritizing ideas, was developed by George Engel in 1977 [**189**]. Helen Longino, in 1990s, emphasized the social nature of knowledge production and contextual empiricism - the role of communal norms and social epistemology in scientific credibility of ideas, and not just individual rationality [**190**]. Funtowicz and Ravetz pioneered Post-Normal Science in 1993, arguing for participatory decision-making when "facts are uncertain, values in dispute, stakes high, and decisions urgent" [**191**]. Sheila Jasanoff, Helga Nowotny, and others extended this thinking into science and technology governance, exploring issues of legitimacy, reflexivity, and public engagement [**192**]. Cartwright emphasised the contextuality and partiality of methods in prioritising ideas [**193**]. All of these approaches can help assess the assumptions and values embedded in prioritisation frameworks; examine the balance between qualitative and quantitative reasoning; address the limits of objectivity and the need for epistemic humility; and navigate ethical trade-offs in global health (e.g., DALYs vs. equity), AI governance, and innovation policy. They are applied in global health, e.g. ethical debates around DALYs, cost-effectiveness, and epistemic justice; technology governance, e.g. responsible research and innovation (RRI), anticipatory governance; research funding, AI and foresight, e.g. meta-analyses of value system, ethics and biases in algorithmic prioritisation. These works continue to influence frameworks such as CHNRI, Delphi, MCDA, and more recently, LLM-based prioritisation tools, by reminding us that the way we evaluate ideas is always shaped by deeper worldviews, institutional interests, and contested values. These four clusters of ideation prioritisation methods - Participatory Budgeting, Citizen Juries, Public Consultations with Weighting, and Cross-Cutting Meta-Theories - illustrate the growing sophistication of participatory and philosophical approaches in shaping what ideas societies pursue. They vary in structure, epistemic assumptions, scalability, and inclusiveness, but all contribute essential tools for navigating complexity, uncertainty, and value pluralism in modern decision-making.

### 3.5.2 Citizen Juries and Deliberative Democracy Forums

Citizen Juries and deliberative forums emerged in the 1970s and 1980s as methods to integrate reasoned public deliberation into policy and research decisions. These approaches were rooted in democratic theory and the ideal of a more informed and reflective public discourse. The Citizen Jury model was pioneered by Ned Crosby, initially in his Ph.D. work at the University of Minnesota and then through his work at the Jefferson Center. The first implementation in 1974 was on healthcare policy [**194**]. In Citizen Juries, a group of randomly selected citizens are given the information and time to discuss a particular issue and make recommendations. They usually bring together 12-24 persons, who receive balanced information, engage with experts, deliberate over multiple days, and ultimately make policy recommendations. This process provides a counterbalance to both elite-driven decision-making and uninformed mass opinion. Parallel efforts in deliberative democracy were advanced by thinkers such as John Dryzek [**195**], who offered philosophical grounding for these practices in their work, shaping the theory behind deliberative forums. The concept of deliberative democracy, which emphasizes the importance of reasoned discussion and debate in decision-making, has roots in ancient Greece. Deliberative Forum is a larger, often one-off, or multi-day event, where diverse lay participants discuss societal challenges with structured facilitation. Crosby's work, particularly with Citizens' Juries, is seen as a practical application of deliberative democratic principles, and Deliberative Forums its extension. Deliberative Poll, as the next step, combines public opinion polling with deliberation, and respondents are surveyed before and after exposure to expert input and discussion. James Fishkin later developed Deliberative Polling® in the 1980s and 1990s as a scalable alternative to small juries, combining opinion polling with structured deliberation to assess shifts in public views after informed discussion [**196, 197**]. He justified Deliberative Polling® as a democratic innovation to balance public opinion and expert knowledge and presented evidence of deliberative polling effects in real-world decision-making. Citizen Juries, Deliberative Forums and Deliberative Polls have been used in various contexts, including policy development, candidate evaluation, and community problem-solving, demonstrating their potential to engage citizens in meaningful ways. These approaches have found wide application in climate adaptation funding, technology ethics, end-of-life care, and AI governance. They exemplify how randomly selected laypersons, when given time, information, and facilitation, can produce thoughtful, legitimate prioritisation of complex issues.

### 3.5.3 Public Consultations and E-Surveys with Weighting

Public consultations and e-surveys with weighting are participatory tools that have matured alongside digital technologies and institutional demands for evidence-informed policymaking. Rather than being the product of a single innovator, these methods have evolved from legal, political, and technological traditions, gaining traction especially from the 1990s onward. The legal foundations for fair consultation are rooted in the legal concept of "legitimate expectation", where the idea of weighting in public consultations were laid by cases such as *Council for Civil Service Unions v Minister for the Civil Service* AC 374 ("GCHQ").

The Gunning Principles, established by Stephen Sedley QC in 1985 in the R v Borough of Brent ex parte Gunning case [**198**], provide a framework for fair and lawful public consultations: timely initiation, transparency of intent, adequate response periods, and serious consideration of inputs. E-surveys, particularly e-mail surveys, emerged in parallel, with early adoption in the 1980s. Kiesler and Sproull published one of the first analyses of email surveys in *Public Opinion Quarterly* in 1986 [**199**]. As internet access expanded, digital surveys became essential tools for gathering structured feedback, with weighting mechanisms introduced to differentiate stakeholder influence based on expertise, role, or representativeness. Pioneering institutions that implemented public consultations include OECD in 1990s and 2000s [**200**], promoting public consultation and participation in governance, and as a formal tool for evidence-based policymaking; then, UK NICE, with integrated consultation and stakeholder weighting in health technology appraisal process in 2004 [**201**]; also, the James Lind Alliance (JLA) [**40,122**] and the Child Health and Nutrition Research Initiative (CHNRI) [**41,42**], which developed equal-weight or weighted e-surveys for research priority-setting; and European Commission, which adopted structured e-consultations in science policy using structured questionnaires with weighted responses [**202**]. Applications range from health research funding (e.g., NIHR, WHO) and policy design (OECD strategies) to science and innovation roadmapping (e.g., Horizon Europe). Tools like Pol.is and IdeaScale exemplify how digital platforms now integrate statistical weighting, multi-criteria decision analysis, and feedback loops into large-scale idea prioritisation exercises.

### 3.5.4 Participatory Budgeting

Participatory Budgeting (PB) is one of the most influential and widely replicated models for citizen-led allocation of public resources. It was first introduced in 1989 in Porto Alegre, Brazil, when the Workers' Party took the office. Developed as a response to growing demands for transparency, equity, and democratic participation, PB allows community members to directly propose, discuss, and vote on how public funds are allocated. The Porto Alegre model emerged during Brazil's transition from dictatorship to democracy, serving as a mechanism to rebuild trust in public institutions and empower marginalized communities. It has been recognized for its success in prioritizing the needs of the poorest areas, improving public services, strengthening governance, and increasing citizen participation. It was a citizen-led process to generate, discuss, and vote on ideas for public funding, redistributing power and enhancing transparency. It influenced deliberative democracy, community-led innovation, and bottom-up research prioritisation frameworks worldwide. Key figures in its development include municipal leaders like Tarso Genro and Raul Pont, along with scholars such as Boaventura de Sousa Santos [**203**], Gianpaolo Baiocchi [**204**], and Brian Wampler [**205**], who later provided robust academic analyses of its mechanisms and outcomes. At its core, PB is structured around a deliberative process: citizens generate ideas, debate them in local assemblies, and vote on proposals, which are then implemented using municipal funds. It has been applied across various domains, such as urban infrastructure, education, health, and youth programs. PB's success in Porto Alegre inspired more than 7,000 replications worldwide, from Paris and New York to Nairobi and Seoul. It has evolved into digital formats, such as Decidim and Consul platforms [**206, 207**], and school-level budgeting initiatives. Its global legacy lies in demonstrating that inclusive, deliberative models can redistribute power, foster equity, and enhance governance legitimacy.

## 3.6 Foundational Philosophical Heuristics and Reasoning Frameworks: Approaches to Prioritising Ideas Beyond Specific Techniques

While many modern prioritisation frameworks rely on structured stakeholder input and formal algorithms, several foundational philosophical heuristics continue to shape how we evaluate, compare, and prioritise competing ideas. These approaches—Occam's Razor, Bayesian Inference, the Dialectical Method, and Epistemic Humility—are not decision-making tools in the narrow sense, but rather enduring intellectual frameworks that guide reasoning under uncertainty, complexity, and value plurality. Their influence spans science, medicine, policy, and artificial intelligence, embedding centuries of epistemological development into contemporary idea evaluation.

### 3.6.1 Occam's Razor

Occam's Razor, also spelled Ockham's Razor, one of the most enduring heuristics in Western intellectual history, prioritises explanatory simplicity: "Prefer simpler ideas when explanations are equal." Although the principle predates him, it was William of Ockham, a 14th-century English Franciscan friar and philosopher, who popularised its application in logic and metaphysics. He articulated the core principle circa 1323–1328 during his major theological and philosophical works, especially in Avignon and Munich. He advocated that "entities should not be multiplied beyond necessity," a phrase later paraphrased to summarise the principle. While Ockham never used the term "Occam's Razor" himself, the idea is evident throughout his major works, particularly *Summa Logicae* (c. 1323) and *Quodlibetal Questions* (c. 1328) [**208,209**]. It introduced foundational principles of logic and parsimony in reasoning, foundational for both evaluating and prioritising ideas based on simplicity and explanatory power. Ockham frequently applied these principles in theological and logical debates. Occam's Razor became integrated into scientific method, logic, AI model selection, medicine, and all disciplines that confront competing hypotheses or explanations. The term "Occam's Razor" was coined retrospectively, gaining currency in the 17th century through philosophers like Johannes Clauberg and Libert Froidmont, and was later formalised by Sir William Hamilton, a renowned Scottish philosopher and professor at the University of Edinburgh, in 1852 [**210**]. The principle, as it is applied today, suggests that among competing hypotheses, the one making the fewest assumptions should be selected, unless additional complexity is clearly warranted. It underpins model selection in scientific research, medical differential diagnostics, and machine learning, where it appears in formalised metrics such as the Akaike Information Criterion (AIC) [**211**] and Bayesian Information Criterion (BIC) [**212**].

### 3.6.2 Bayesian Inference

Bayesian inference provides a formal mechanism for updating beliefs about hypotheses or models in light of new evidence. Though named after Thomas Bayes, English Presbyterian minister and mathematician, who outlined a specific probabilistic problem in his 1763 posthumous essay, the method was given modern mathematical form by Pierre-Simon Laplace in the late 18th and early 19th centuries. Bayes' original work, *An Essay Towards Solving a Problem in the Doctrine of Chances*, was edited and published by his friend Richard

Price in the *Philosophical Transactions of the Royal Society* [**213**]. This publication introduced what we now call Bayes' Theorem, which allows updating the probability of a hypothesis in light of new evidence. Laplace, French polymath who influenced the fields of physics, astronomy, mathematics, engineering, statistics and philosophy, extended Bayes' ideas into a general framework for probabilistic reasoning, applying it to astronomy, population statistics, and political decision-making [**214**]. Bayesian inference calculates the probability of a hypothesis (posterior) by combining its prior probability with the likelihood of observing current evidence. It inherently incorporates a penalty for overly complex hypotheses, which is a formal echo of Occam's Razor. Importantly, it can be used to rank competing ideas based on both plausibility and fit with data. Today, it underlies model selection, decision theory, policy forecasting, health technology assessment, AI safety frameworks, and many more areas of progress. Its conceptual elegance lies in combining subjective beliefs with objective evidence through formal probability.

### 3.6.3 Dialectical Method – Thesis, Antithesis, Synthesis

The dialectical method—resolving tension between opposing ideas through synthesis—originated in ancient philosophical debates but was systematised in modern form by German idealists, particularly Johann Gottlieb Fichte and Georg Wilhelm Friedrich Hegel. Though Hegel did not use the terms "thesis–antithesis–synthesis" explicitly, this triadic model was later used to describe the dynamic movement of ideas in his philosophy, especially in his book *Phenomenology of Spirit* in 1807 [**215**]. The method involves an initial idea (thesis), its negation or contradiction (antithesis), and their resolution in a higher-order synthesis, which usually incorporates the elements of both. This process is not merely cyclical but developmental: each synthesis becomes a new thesis, advancing understanding through iterative contradiction and integration. Heinrich Moritz Chalybäus was the first to formalise this triadic structure in 1837, retroactively applying it to Hegel's work [**216**]. The dialectical method influenced prioritisation of ideas in critical theory, scenario planning, and design thinking, particularly wherever synthesis of conflicting perspectives is essential. While it wasn't originally formulated for prioritising ideas in a modern evaluative sense, it has been used for evaluating competing perspectives, identifying contradictions, and advancing knowledge—making it relevant to meta-theoretical prioritisation in philosophy, politics, and science.

### 3.6.4 Epistemic Humility and Pluralism – Valuing Diverse Viewpoints in Evaluation

The principles of epistemic humility and pluralism challenge the idea that knowledge is objective, singular, or complete. Rooted in ancient Greek philosophy, particularly in Socratic dialogue, they found renewed expression in modern political philosophy, feminist epistemology, decolonial theory, and science and technology studies. Socrates' assertion of his own ignorance (*Plato's Apology*) provided a foundation for epistemic humility [**217**]. In the 19th century, John Stuart Mill argued in 1859, in his book *On Liberty*, that exposure to dissenting views strengthens knowledge [**218**]. In the 20th century, Paul Feyerabend's *Against Method* (1975) rejected rigid scientific orthodoxy, advocating instead for methodological pluralism and epistemological anarchism: "Let many flowers bloom" [**202**]. Further contributions came from many more recent thinkers, who all argued that diverse perspectives, including those from lay, indigenous, and marginalised groups, improve the legitimacy, robustness, and ethical grounding of idea evaluation. Today, these principles are embedded in participatory research models like CHNRI's weighting and setting thresholds by diverse shareholders, or James Lind Alliance partnerships. The heuristics and meta-frameworks explored in this section, which included parsimony, probabilistic updating, dialectical synthesis, and epistemic humility, offer essential cognitive and ethical guidance in the face of competing ideas. Though they emerged from different epochs and disciplines, they converge on a common premise: the process of prioritisation is not merely technical, but also philosophical. By combining these timeless principles with contemporary participatory methods, we can foster more reasoned, inclusive, and reflexive approaches to idea evaluation in science, policy, and innovation.

**Discussion**

**Toward a Unified Science of Ideas**

The exercise of systematically identifying and structuring the methods that were historically used by humans to generate, evaluate, and prioritise ideas has revealed several insights. Despite their disciplinary diversity, these methods exhibit striking structural, functional, and philosophical parallels. From ancient heuristics to machine learning algorithms, from tribal councils to digital deliberative platforms, the fundamental aim has remained the same: how to make sense of competing possibilities and decide which ideas are worth acting upon.

When viewed through an ideometric lens, several recurring patterns emerge across methods, regardless of their disciplinary origin, epistemic orientation, or technical complexity, uncovering interesting parallels across disparate traditions. A number of them, especially the more contemporary ones, seem to implicitly or explicitly follow a layered process that mirrors the stages described in the brain's "sense of ideas" concept: generation, evaluation, and prioritisation of ideas {**1**}. For example, brainstorming generates divergent options, Delphi panels evaluate them through expert feedback, and AHP ranks them using quantitative pairwise comparisons. The CHNRI method follows a similar triad: generation through open solicitation, evaluation using expert scoring, and prioritisation by aggregated weighted scores.

There is also a persistent tension across methods. It lies in balancing subjective human judgment with the aspiration for objectivity. Peer review, Delphi, and NGT depend on subjective expert or stakeholder input, while TRLs, citation metrics, and MCDA aim to quantify value. AI-based systems like LLMs and reinforcement learning introduce a new variant: machine-generated outputs grounded in massive amounts of data, although they are trained on inherently subjective human content. This interplay suggests that ideometric tools should not aspire to remove subjectivity. Instead, they need to structure and balance it transparently.

Recursive refinement is another common theme. Many techniques rely on iterative feedback loops. Delphi rounds, agile sprints, design thinking cycles, and machine learning training epochs all reflect recursive logic. They aim not for immediate certainty, but for progressive approximation. This emphasis on refinement over resolution implies that ideometric systems are most effective when treated as dynamic, living processes rather than static decision tools.

Across ideometric methods, the act of aggregation seems to be the "core engine". Whether scores, votes, arguments, or preferences are in question, aggregation is the pivotal mechanism that transforms individual judgments into collective decisions. MCDA aggregates weighted scores; prediction markets aggregate probabilistic bets; knowledge graphs aggregate semantic relations. This central role of aggregation suggests that future ideometric tools should continue to innovate around how inputs are synthesized, how transparency is preserved, and how differing values are represented.

Finally, each method encodes implicit assumptions about what counts as a "good" idea, based on some embedded values and norms: whether it is logical coherence (e.g., deductive reasoning), novelty (e.g., patent originality), societal need (e.g., JLA PSPs), stakeholder consensus (e.g., Deliberative Democracy), or impact on equity (e.g., CHNRI). Philosophical heuristics such as Occam's Razor or epistemic humility make these values explicit, while technical tools like AHP or CHNRI embed them in scoring rubrics. Recognizing these normative assumptions is crucial, because ideometrics is never "just technical". It needs to be appreciated that it is ethical, can be subjective, and is political, too.

**What Is Still Missing?**

On the note of the role of politics in this field of science, as we already indicated in the introduction, our current work surely placed insufficient attention to the role of political authority and financial power on prioritising ideas. It is difficult to scientifically address political decisions that seem to conflict rational reasoning but still have large impact on the ideas that are prioritised. This is especially true when such political decisions have wide public support based on either insufficient information, or even disinformation [**7**]. Furthermore, funding decisions on future strategy by the main funding agencies can also be highly political and thus have impact on prioritisation of ideas for the entire scientific community. Michel Foucault and several other great thinkers through history have addressed these themes profoundly in their highly influential work [**10**], so we saw it as beyond the scope of our current efforts to define and position Ideometrics as a scientific field.

Then, we already explained in the introductory section that we were unable to grasp all the approaches to generating, evaluating and prioritising ideas that may exist. There are certainly approaches to priority setting that may be quite well known, but we managed to miss them for a number of reasons. Language barrier might be one – we couldn't really assess the tools that likely exist in many local settings world-wide and are described in the literature in local languages.

Another substantial limitation is that our work approached this field from a largely western and rationalist perspective. We mentioned in the opening paragraphs of this work that, in more traditional cultures, the emphasis in choosing best ideas could often be based on ancestral traditions and folklore, the experiences of elders, intuition, revelation and spiritual enlightenment [**3,4**]. We did not attempt to enter that vast space and try to highlight the most enduring and/or promising approaches.

It is possible that, at times, our work could give impression that it is mixing approaches that are too different in their specificity, from the rather sublime debate that characterised the ancient Greece and its philosophers to the prescriptive structures of the CHNRI or JLA methods. In trying to be systematic in a large task that we undertook, we left together the methods that were at disposal thousands of years ago and the modern, AI-based tools, although times and options have surely changed over that period.

There is also an area within ideometrics where evidence-informed policymaking needs to be based on the information that is emerging in real time, through iterative rounds, as we saw in the COVID-19 pandemic [**219,220**]. Within this specific context, the evolution of so-called "rapid reviews" and "living reviews" [**221,222**], and the role of systematic reviews in general, as a methodologically sophisticated approach to summarising the empirical evidence-base [**223**], deserve special attention within the field of Ideometrics.

In priority-setting processes that involve many individuals, conflicts of interest (COIs) can significantly influence outcomes, potentially distorting the collective judgment. Individuals with financial, professional, or personal stakes in particular outcomes may consciously or unconsciously advocate for priorities that align with their interests rather than the broader public good. When such conflicts are not disclosed or adequately managed, they can bias the ranking of research topics, divert resources away from areas of greatest need, and undermine the credibility of the entire process. This risk increases in multi-stakeholder exercises where participants come from diverse sectors, including academia, industry, government, and civil society. To safeguard the integrity and legitimacy of priority-setting exercises, it is essential to require transparent disclosure of all potential COIs, adopt independent facilitation, and apply structured consensus-building methods that dilute individual influence and promote collective reasoning. Clear governance mechanisms are key to ensuring fairness, accountability, and trust in the final outcomes [**224**].

There is another area well-worth exploring and developing within Ideometrics: the one of empirical assessment whether certain ideas are indeed "better" in practice. Namely, researchers can track real-world implementation and outcomes of competing ideas over time. Comparative evaluations, such as randomized controlled trials, natural experiments, or pilot programs, can measure the effectiveness, cost-efficiency, and scalability of ideas in different contexts. Outcome indicators, like health impact, uptake, economic return, or social acceptance, provide measurable benchmarks. Longitudinal studies and impact evaluations can further reveal which ideas deliver sustained benefits. By linking idea selection to empirical feedback loops, we can iteratively refine priority-setting processes and identify which types of ideas consistently lead to meaningful, evidence-backed improvements [**225**].

Finally, it has to be fairly acknowledged that some of the best ideas throughout human history, that eventually resulted in large positive impact on societies, science, technology, art, and human experience in general, came from a sequence of serendipitous events that happened seemingly by mere chance, in a very beneficial way. In fact, there were also great ideas that came as a result of a gross error in rational judgement, or through entirely disinformed sequence of thought or actions, but resulted in a very positive impact [**226**]. That may be the ultimate challenge for Ideometrics, as a field of science, to learn how to harness without unacceptable risks.

**From Framework to Field: Developing Ideometrics as a Science**

Having identified, organised, and synthesised a broad range of methods to support Ideometrics, we now turn to the question: how to elevate *Ideometrics* from a useful framework to a recognised field of scientific inquiry? The first step is empirical. Ideometrics must subject itself to the very tools it has catalogued. We therefore propose to conduct a quantitative bibliometric analysis of the "footprint" of each ideometric method in the scholarly literature. Using established citation databases such as Scopus, Web of Science, and Google Scholar, we aim to track the total number of publications referencing each method; measure citation velocity and longevity (e.g., half-life); assess disciplinary spread and cross-sectoral uptake; identify highly cited papers and thought leaders in each tradition; evaluate their adoption in policy, health, engineering, business, and education. This analysis will not only provide an empirical baseline for future research, but also reveal which methods have achieved practical traction, which remain underutilised, and which have declined in relevance.

In parallel, we aim to introduce a set of Reporting Guidelines for Ideometric Exercises – i.e. attempts to generate, evaluate and prioritise ideas using any of the methods - analogous to the PRISMA or CONSORT statements in health research [**227**,**228**]. These guidelines would standardise the documentation of ideation exercises, covering elements such as: (i) who participated, and how they were selected; what criteria were used for evaluation and why; which aggregation methods were employed; how transparency, replicability, and equity were ensured; and how results are intended to inform policy or decision-making. These guidelines would promote comparability across methods, improve the reproducibility of ideometric studies, and open the door to meta-analyses of approaches that addressed the same challenge, and enable comparisons across domains (e.g., by comparing Delphi vs. CHNRI vs. LLM rankings on the same problem set). They could also form the basis of journal submission requirements for ideometric-based research papers.

**A Research Agenda for the Future of Ideometrics**

To establish ideometrics as a coherent and dynamic field of science, we propose the following research and development priorities: (i) Building a living repository of methods and use cases: we propose the creation of a curated, openly accessible digital repository that catalogues ideometric methods, case studies, scoring templates, decision matrices, and user experiences. This could become a "Wikipedia meets GitHub" for idea management—both archival and interactive. (ii) Conduct Cross-Method Comparative Trials: Just as new treatments are tested against standards of care, ideometric methods should be empirically compared. For example: if CHNRI, Delphi, AHP, and an LLM-based method are all applied to the same set of global health research questions, do they produce similar priorities? If not, why? What assumptions are driving the divergence? Comparative trials would reveal method sensitivity, value alignment, and practical feasibility. (iii) Investigate Cognitive and Social Dynamics: As ideometrics increasingly shapes real-world decisions, we must understand its psychological and sociological dimensions. How do different stakeholder groups engage with ideometric tools? What cognitive, social, financial, political, power-driven, and other biases persist despite formal structures? How do power dynamics influence prioritisation in participatory settings? This work would draw from behavioural science, sociology of science, and ethics. (iv) Integrate AI and Human-Centred Design: Our next frontier is the development of a user-friendly, transparent, and democratic software tool for ideometrics-based decision support. We are currently working on such a platform. It will allow users (individuals, organisations, governments) to input ideas and associated metadata; facilitate transparent scoring across selected criteria; include participatory features (public voting, weighting, deliberation); offer AI-enhanced clustering, validation, and prioritisation tools; and provide structured outputs that are traceable, auditable, and replicable. This software will serve as both a research engine and a policy tool, bridging the gap between academic rigour and practical application. Its ultimate aim is to support evidence-informed, participatory, and adaptive decision-making across sectors.

**Practical Implementation and Societal Impact**

For ideometrics to realise its transformative potential, it must go beyond academic theory. We envision its practical implementation across several domains: (i) *science and research*: national research councils, funders, and global health agencies can adopt ideometric frameworks to inform priority-setting, peer review reform, and grant allocation. Tools like CHNRI, GRADE, JLA and many other methods have already shown promise, while newer AI-assisted approaches will scale these to new domains. (ii) *public policy*: governments can integrate ideometric tools into strategic foresight, infrastructure planning, and citizen engagement. Participatory budgeting, e-surveys with weighting, and deliberative forums—now digitally enhanced—can align public values with policy design. (iii) *corporate strategy and innovation:* Businesses and startups can use ideometric dashboards to triage product ideas, assess R&D pipelines, and align investments with market trends. Frameworks like AHP, Stage-Gate, and Real Options Analysis can now be implemented more intuitively via intelligent software. (iv) *education and capacity building:* Ideometrics can be taught in universities, professional development programmes, and leadership training. Just as students learn experimental design or financial modelling, future leaders should understand how to think critically and systematically about ideas. (v) *global governance and foresight:* In addressing transnational challenges—climate change, pandemics, AI regulation—international institutions must weigh competing solutions under uncertainty. Ideometric tools that combine inclusiveness with analytical rigour can support transparent, democratic, and equitable global coordination.

**Conclusion: A New Scientific Imagination**

Working on this paper, we increasingly felt like archaeologists who have discovered a science that has been here all along for centuries, present in virtually every line of human activity, but that no-one could easily notice, let alone grasp in its wholeness, simply because it was dispersed too widely and broadly, in chunks too small and diverse to connect. Connecting it into a mosaic of "Ideometrics" across many fields, using the simple framework to "generate-evaluate-prioritise ideas", it felt like an effort in archaeology of human approaches to ideation, assessment of the new ideas and choosing those that seemed worth following to people over times and spaces.

The science of ideometrics does not seek to mechanise human creativity or reduce thought to computation. Rather, it invites us to see the life of ideas, from their inception to their selection, as a subject of structured inquiry. It respects the depth of ancient heuristics and the promise of modern algorithms. It embraces pluralism, transparency, and methodological humility. It recognises that ideas are not only cognitive artefacts but also social commitments: they are shaped by who speaks, who listens, and how we all make decisions. By documenting its foundations, mapping its landscape, and outlining a roadmap for growth, we will help this work to catalyse the emergence of ideometrics as a vibrant interdisciplinary field. Its potential is not only intellectual, but also highly practical: a better science of ideas can lead to better ideas, and better ideas can, in turn, lead to a better world.

**Acknowledgements**

This work did not receive any funding support. IR wrote the first draft of this work. AS then provided substantial critical feedback and guidance on how to improve and position this work within the global intellectual context and facilitate its future implementation in policy and practice. IR wishes to thank his wife Tonkica for continuing intellectual stimulation that assists him in conceptualising and drafting his creative work.

**REFERENCES**

1. Rudan I. Editor's view: Is the brain's perception of ideas an underappreciated human sense? J Glob Health. 2024 Dec 6;14:01002.
2. Lovejoy AO. Reflections on the history of ideas. J History Ideas 1940; 1:3-23
3. Morley N. Approaching the Ancient World: Theories, Models and Concepts in Ancient History. Routledge: London, 2004.
4. Morin O. How Traditions Live and Die. Oxford University Press, Oxford, 2016.
5. Bibliometrics: https://en.wikipedia.org/wiki/Bibliometrics; Accessed: 27 Jul 2025.
6. Econometrics: https://en.wikipedia.org/wiki/Econometrics; Accessed: 27 Jul 2025.
7. Rudan I. Editor's view: Value of information in the 21st century - examples from science, medicine, policy, media, and markets. J Glob Health. 2025;15:01003.
8. Rudan I. Editor's view: What makes science successful? J Glob Health. 2025 Jul 21;15:01005.
9. Censorship. Available from: https://en.wikipedia.org/wiki/Censorship; Accessed: 27 Jul 2025.
10. Foucault M. Power/Knowledge. Selected Interviews and Other Writings. Pantheon Books, New York, 1972.
11. Heuristic. Available from: https://en.wikipedia.org/wiki/Heuristic; Accessed: 27 Jul 2025.
12. Humanities, Arts and Social Sciences. Available from: https://en.wikipedia.org/wiki/Humanities,_arts,_and_social_sciences; Accessed: 27 Jul 2025.
13. Quantitative Research. Available from: https://en.wikipedia.org/wiki/Quantitative_research; Accessed: 27 Jul 2025.
14. James W. The Principles of Psychology, vol. 1 and 2. New York: Henry Holt and Co, 1890.
15. Dujardin E. Les Lauriers sont coupés. Librairie de la Revue Indépendante, Paris, 1887.
16. Joyce J. Ulysses. Shakespeare and Company, Paris, 1922.
17. Elbow P. Writing without Teachers. Oxford University Press, New York, NY, 1973.

18. Altshuller GS, Shapiro RV. O psikhologii izobretatelskogo tvorchestva (*On the psychology of inventive creation*). Voprosy psikhologii 1956; 6:37-49.
19. Altshuller GS. Creativity as an Exact Science: The Theory of the Solution of Inventive Problems. CRC Press, London, 1984.
20. Zwicky F. Morphological Astronomy. Springer-Verlag, Berlin, 1957.
21. Zwicky F. Discovery, invention, research through the morphological approach. Macmillan, Toronto, 1969.
22. de Bono, E. The Use of Lateral Thinking. Jonathan Cape, London, 1967.
23. Eberle RF. SCAMPER: Games for Imagination Development. D.O.K. Publishers, Buffalo, NY, 1971.
24. Buzan T. Use your head. BBC Books, London, 1974.
25. Buzan T, Buzan B. The mind mapping book; radiant thinking- the major evolution in human thought. BBC Books, London, 1993.
26. Tversky A, Kahneman D. Judgment under uncertainty: Heuristics and biases. Science. 1974;185(4157):1124-31.
27. Kahneman D. Thinking, fast and slow. Farrar, Straus and Giroux, New York, NY, 2011.
28. de Bono Edward. Six Thinking Hats. Penguin Books, London, 1985.
29. Osborn AF. How to Think Up. McGraw-Hill Book Co, New York, NY, 1942.
30. Osborn AF. Applied imagination: Principles and procedures of creative problem solving. Charles Scribner's Sons, New York, NY, 1953.
31. Merton RK, Fiske M, Kendall PL. The focused interview; a manual of problems and procedures. Free Press, Glencoe, IL, 1956.
32. Dichter E. The Strategy of Desire. Doubleday, New York, NY, 1960
33. Morgan DL. Focus Groups as Qualitative Research. 2nd ed. Vol. 16. Qualitative Research Methods Series. Sage Publications, Thousand Oaks, CA,1997.
34. Dalkey N, Helmer O. An Experimental Application of the DELPHI Method to the Use of Experts. Management Science 1963; 9:458-467.
35. Delbecq AL, Van de Ven AH. A Group Process Model for Problem Identification and Program Planning. J Appl Behav Sci 1971; 7:466-492.
36. Delbecq AL, Van de Ven AH, Gustafson DH. Group Techniques for Program Planning: A Guide to Nominal Group and Delphi Processes. Scott, Foresman, Glenview, IL,1975.
37. Brown J, Isaacs D. The World Café: Shaping our futures through conversations that matter. Berrett-Koehler, San Francisco, 2005.
38. Owen H. Open Space Technology: A user's guide. Berrett-Koehler, San Francisco, 1997.
39. Bingham A, Spradlin D. The Open Innovation Marketplace: Creating value in the challenge driven enterprise. FT Press, Upper Saddle River, NJ, 2011.
40. Partridge N, Scadding J. The James Lind Alliance: patients and clinicians should jointly identify their priorities for clinical trials. Lancet 2004; 364:1923-4.
41. Rudan I, El Arifeen S, Black RE. A Systematic Methodology for Setting Priorities in Child Health Research Investments. In: Child Health and Nutrition Research initiative (CHNRI): A New Approach for Systematic Priority Setting In Child Health Research Investment. Dynamic Printers, Dhaka, 2006.
42. Rudan I, Gibson JL, Ameratunga S, El Arifeen S, Bhutta ZA, Black M, Black RE, Brown KH, Campbell H, Carneiro I, Chan KY, Chandramohan D, Chopra M, Cousens S, Darmstadt GL, Meeks Gardner J, Hess SY, Hyder AA, Kapiriri L, Kosek M, Lanata CF, Lansang MA, Lawn J, Tomlinson M, Tsai AC, Webster J; Child Health and Nutrition Research Initiative. Setting priorities in global child health research investments: guidelines for implementation of CHNRI method. Croat Med J. 2008;49(6):720-33.
43. Anonymous. Crowdsourcing Launch for Survey Analytics. MRweb, 30 Sep 2008. Available from: https://www.mrweb.com/drno/news8955.htm; Accessed: 27 Jul 2025;
44. Ideascale. Available from: https://ideascale.com/about-ideascale/; Accessed: 27 Jul 2025;
45. Human-Centered Design. Available from: https://en.wikipedia.org/wiki/Human-centered_design; Accessed: 27 Jul 2025;
46. Norman DA. The Psychology of Everyday Things. Basic Books, New York, NY, 1988.
47. IDEO.org. The Field Guide to Human-Centered Design. IDEO.org / Design Kit, San Francisco, CA, 2009.
48. Simon HA. The sciences of the artificial. MIT Press, Cambridge, MA, 1969.
49. McKim RH. Experiences in visual thinking. Brooks/Cole, Monterey, CA, 1973.
50. Brown T. Change by Design. HarperBusiness, New York, NY, 2009.
51. Schwaber K, Sutherland J. SCRUM Development Process. OOPSLA '95 conference, Austin, Texas, 1995.
52. Beck K, Beedle M, van Bennekum A, Cockburn A, Cunningham W, Fowler M, Grenning J, Highsmith J, Hunt A, Jeffries R, Kern J, Marick B, Martin R, Mellor S, Schwaber K, Sutherland J, Thomas D. Manifesto for Agile Software Development. Agile Alliance, 2001. Available from: https://agilemanifesto.org/ (Accessed: 27 Jul 2025)
53. Knapp J, Zeratsky J, Kowitz B. Sprint. London, Bantam Press, 2016.
54. Hackaton. Available from: https://en.wikipedia.org/wiki/Hackathon; Accessed: 27 Jul 2025;
55. Briscoe G, Mulligan C. Digital Innovation: The Hackathon Phenomenon. 2014. Available from: https://qmro.qmul.ac.uk/xmlui/handle/123456789/11418; Accessed: 27 Jul 2025;
56. Irani L. Hackathons and the making of entrepreneurial citizenship. Science, Technology and Human Values 2015; 40: 799-824.
57. Lodato T, DiSalvo C. Issue-oriented Hackathons as material participation. New Media and Society 2016; 18: 539–557.
58. Blank SG. The Four Steps to the Epiphany: Successful Strategies for Products that Win. Quad/Graphics, Sussex, WI, 2005.
59. Ries, E. The Lean Startup: How Today's Entrepreneurs Use Continuous Innovation to Create Radically Successful Businesses. Crown Business, New York, NY, 2011.
60. Holland JH. Adaptation in Natural and Artificial Systems. University of Michigan Press, Ann Arbor, 1975.
61. Swanson DR. Fish oil, Raynaud's syndrome, and undiscovered public knowledge. Perspect Biol Med. 1986; 30:7-18.
62. Smalheiser NR, Swanson DR. Using ARROWSMITH: a computer-assisted approach to formulating and assessing scientific hypotheses. Comput Methods Programs Biomed. 1998; 57:149-53.
63. Pyysalo S, Baker S, Ali I, Haselwimmer S, Shah T, Young A, Guo Y, Högberg J, Stenius U, Narita M, Korhonen A. LION LBD: a literature-based discovery system for cancer biology. Bioinformatics. 2019; 35:1553-1561.
64. IBM Watson Discovery. Available from: https://www.ibm.com/products/watson-discovery; Accessed: 27 Jul 2025.
65. Semantic Scholar. Available from: https://www.semanticscholar.org/; Accessed: 27 Jul 2025;
66. Goodfellow IJ, Pouget-Abadie J, Mirza M, Xu B, Warde-Farley D, Ozair S, Courville A, Bengio Y. Generative Adversarial Nets. Advances in Neural Information Processing Systems 27: 2672-2680, 2014.
67. Vaswani A, Shazeer N, Parmar N, Uszkoreit J, Jones L, Gomez AN, Kaiser L, Polosukhin I. Attention Is All You Need. 2017. https://arxiv.org/abs/1706.03762.
68. Radford A, Wu J, Child R, Luan D, Amodei D, Sutskever I. Language Models are Unsupervised Multitask Learners. OpenAI Technical Report on GPT-2, OpenAI, 2019.
69. Brown TB, Mann B, Ryder N, Subbiah M, Kaplan JD, Dhariwal P, et al. Language Models are Few-Shot Learners. Part of Advances in Neural Information Processing Systems 33, NeurIPS, 2020. Available from: https://arxiv.org/abs/2005.14165; Accessed: 27 Jul 2025.
70. Scholarly Peer Review. Available from: https://en.wikipedia.org/wiki/Scholarly_peer_review; Accessed: 27 Jul 2025.
71. Burnham JC. The Evolution of Editorial Peer Review. JAMA. 1990;263(10):1323–1329.
72. Institute of Medicine Committee to Improve the National Institutes of Health Consensus Development Program. Consensus Development at the NIH: Improving the Program. National Academies Press (US), Washington, DC, 1990.
73. Clancy CM, Slutsky J. AHRQ's Evolving Role in Comparative Effectiveness Research. Health Affairs. 2007;26(Suppl 1):w140-w145.
74. Saaty TL. A scaling method for priorities in hierarchical structures. Journal of Mathematical Psychology, 15: 234–281, 1977.
75. Saaty TL. The Analytic Hierarchy Process: Planning, Priority Setting, Resource Allocation. McGraw-Hill, New York, 1980.
76. Jiang W, Marggraf R. The origin of cost-benefit analysis: a comparative view of France and the United States. Cost Eff Resour Alloc. 2021;19:74.
77. Marshall A. Principles of Economics. Macmillan, London, 1890.
78. Eckstein O. Water-Resource Development: The Economics of Project Evaluation. Harvard University Press, Cambridge, MA, 1958.
79. Mishan EJ. Cost-Benefit Analysis: An Introduction. George Allen & Unwin, London, 1971.
80. Weisbrod BA. Economics of Public Health: Measuring the Impact of Diseases. 2nd Edition, University of Pennsylvania Press, Philadelphia, PA, 1961.
81. Weinstein MC, Stason WB. Foundations of cost-effectiveness analysis for health and medical practices. N Engl J Med 1977;296:716-21.
82. Fisher I. The rate of interest: Its nature, determination and relation to economic phenomena. Macmillan, New York, 1907.

83. Fisher I. The Theory of Interest. Macmillan, New York, 1930.
84. Dean J. Capital budgeting: Top management policy on plant, equipment, and product development. Columbia University Press, New York, 1951.
85. Eugen von Böhm-Bawerk, Positive Theorie des Kapitales. Wagner, Innsbruck, 1889.
86. Keynes JM. The General Theory of Employment, Interest and Money. Macmillan, London, 1936.
87. Garfield E. Citation indexes for science; a new dimension in documentation through association of ideas. Science. 1955;122:108-11.
88. Jaffe AB, Trajtenberg M, Henderson R. Geographic localization of knowledge spillovers as evidenced by patent citations. Quart J Economics, 108: 577–598, 1993.
89. Hall BH, Jaffe AB, Trajtenberg M. The NBER Patent Citations Data File: Lessons, Insights and Methodological Tools. NBER Working Paper No. 8498, 2001. https://doi.org/10.3386/w8498
90. Garfield E. Citation analysis as a tool in journal evaluation. Science. 1972;178:471-9.
91. Hirsch JE. An index to quantify an individual's scientific research output. Proc Natl Acad Sci U S A. 2005;102(46):16569-72.
92. DORA: San Francisco Declaration on Research Assessment. Available from: https://sfdora.org/read/; Accessed: 27 Jul 2025.
93. Héder M. From NASA to EU: the evolution of the TRL scale in Public Sector Innovation. Innovation Journal 2017; 22(2):1.
94. Mankins JC. Technology Readiness Levels: A White Paper. Advanced Concepts Office, Office of Space Access and Technology, NASA, 1995.
95. Mankins JC. Technology Readiness Assessments: A Retrospective. Acta Astronautica, 65: 1216–1223, 2009.
96. Franklin B. Letter to Joseph Priestley, September 19, 1772.
97. Kuhn HW, Tucker AW. Nonlinear programming. In J. Neyman (Ed.), Proceedings of the Second Berkeley Symposium on Mathematical Statistics and Probability (pp. 481-492). University of California Press, Berkeley, 1951.
98. Raiffa H, Schlaifer R. Applied Statistical Decision Theory. Division of Research, Harvard Business School, 1961.
99. Roy B. Classement et choix en présence de points de vue multiples. Revue Française d'Informatique et de Recherche Opérationnelle, 2(V1): 57-75, 1968.
100. Keeney RL, Raiffa H. Decisions with Multiple Objectives: Preferences and Value Tradeoffs. John Wiley and Sons, New York, 1976.
101. Zionts S. MCDM---If Not a Roman Numeral, Then What? Interfaces 9(4):94-101, 1979.
102. Zionts S, Wallenius J. An Interactive Multiple Objective Linear Programming Method for a Class of Underlying Nonlinear Utility Functions. Management Science 1983; 29:519-529.
103. https://en.wikipedia.org/wiki/Multiple-criteria_decision_analysis; Accessed: 27 Jul 2025;
104. https://en.wikipedia.org/wiki/SWOT_analysis; Accessed: 27 Jul 2025.
105. Humphrey, A. SWOT Analysis for Management Consulting. SRI Alumni Newsletter, SRI International, December 2005, pp. 7-8.
106. Learned EP, Christensen CR, Andrews KR, Guth WD. Business Policy: Text and Cases. R. D. Irwin, Homewood, IL, 1965.
107. Churchman CW, Ackoff RL, Arnoff EL. Introduction to Operations Research. New York: John Wiley & Sons, 1957.
108. Boehm BW. Software Engineering Economics. Prentice-Hall, Englewood Cliffs, NJ, 1981.
109. Belton V, Stewart TJ. Multiple criteria decision analysis: an integrated approach. Kluwer Academic Publishers, Boston, 2002.
110. Pugh S. Total Design: Integrated Methods for Successful Product Engineering. Addison-Wesley, Wokingham, UK, 1990.
111. Fuller, R. Buckminster. Synergetics: Explorations in the Geometry of Thinking. Macmillan, London, 1975,
112. Drucker P. Management Challenges for 21st Century. Harper Business, New York, NY, 1999.
113. Martin RL. The Design of Business: Why Design Thinking is the Next Competitive Advantage. Harvard Business School Publishing, Cambridge, MA, 2009.
114. Surowiecki J. The Wisdom of Crowds: Why the Many Are Smarter Than the Few. Doubleday, New York, 2004.
115. Galton F. Vox Populi. Nature 1907; 75: 450–451.
116. Saldivar J, Daniel F, Cernuzzi L, Casati F Online idea management for civic engagement: A study on the benefits of integration with social networking. ACM Transactions on Social Computing 2019); 2:1-29.

117. Hanson R. Could gambling save science? Encouraging an honest consensus. Social Epistemology 1990; 4: 227–232.
118. Hanson R. Combinatorial information market design. Information Systems Frontiers 2003; 5: 105–119.
119. Iowa Electronic Markets. Available from: https://en.wikipedia.org/wiki/Iowa_Electronic_Markets; Accessed: 27 Jul 2025;
120. Wolfers J, Zitzewitz E. Prediction Markets. Journal of Economic Perspectives 2004; 18:107–126.
121. Prediction Market. Available from: https://en.wikipedia.org/wiki/Prediction_market; Accessed: 27 Jul 2025.
122. James Lind Alliance Guidebook. Available from: https://www.jla.nihr.ac.uk/jla-guidebook/; Accessed: 27 Jul 2025.
123. Cowan K. The James Lind alliance: tackling treatment uncertainties together. J Ambul Care Manage. 2010;33:241-8.
124. Rudan I. Setting health research priorities using the CHNRI method: IV. Key conceptual advances. J Glob Health. 2016; 6:010501.
125. Rudan I, Yoshida S, Wazny K, Cousens S (Eds): Measuring ideas: The CHNRI method. A solution for setting research priorities. International Society of Global Health, Edinburgh, 2022.
126. Shuen A. Web 2.0: A Strategy Guide. O'Reilly Media, Sebastopol, CA, 2008.
127. Paine KD. Measure What Matters: Online Tools for Understanding Customers, Social Media, Engagement, and Key Relationships. John Wiley and Sons, Hoboken, NJ, 2011.
128. Priem J, Taraborelli D, Groth P, Neylon C. Altmetrics: A Manifesto. 2010. Originally published at: http://altmetrics.org/manifesto; Available from: https://digitalcommons.unl.edu/scholcom/185/; Accessed: 27 Jul 2025.
129. Berger J, Milkman KL. What makes online content go viral? J Marketing Res 2012; 49:192–205.
130. Campbell IH, Rudan I. Effective approaches to public engagement with global health topics. J Glob Health 2020;10:01040901.
131. Campbell IH, Rudan I. Analysis of public engagement with ten major global health topics on a social network profile and a newspaper website. J Glob Health 2020; 10:010902.
132. Colicev A, Malshe A, Pauwels K, O'Connor P. Improving consumer mindset metrics and shareholder value through social media: The different roles of owned and earned media. J Marketing 2018; 82: 37-56.
133. Aristotle. Prior Analytics. Around 350 BC. Classical archive, Massachusetts Institute of Technology. Available from: https://classics.mit.edu/Aristotle/prior.html; Accessed: 27 Jul 2025.
134. Aristotle. Metaphysics (Book IV). Around 350 BC. Available from: https://ia802806.us.archive.org/30/items/aristotlesmetaph0001aris/aristotlesmetaph0001aris.pdf; Accessed: 27 Jul 2025.
135. Bacon F. Novum Organum 1620. Montague B (editor and translator), volume I-III. Parry and MacMillan, Philadelphia, PA, 1854.
136. Hume D. A Treatise of Human Nature 1739. Nidditch PH (editor). Clarendon Press, Oxford, 2009.
137. Bernard C. An Introduction to the Study of Experimental Medicine 1865. (English translation). Macmillan & Co, New York, NY, 1927.
138. Mill JS. A System of Logic. J. W. Parker, London, 1843.
139. Popper KR. Logik der Forschung. Springer, Wien, 1934.
140. Kuhn TS. The Structure of Scientific Revolutions. University of Chicago Press, Chicago, IL, 1962.
141. Thurstone LL. A law of comparative judgment. Psychological Review, 1927. 34:273-286.
142. Bradley RA, Terry ME. Rank analysis of incomplete block designs: I. The method of paired comparisons. Biometrika, 1952; 39:324-345.
143. Elo AE. The Rating of Chessplayers, Past and Present. Arco, New York, NY, 1978.
144. Bens I. Facilitating with Ease! Jossey-Bass / Pfeiffer, Hoboken, NJ, 2000.
145. Brassard M, Ritter D. The Memory Jogger II: A Pocket Guide of Tools for Continuous Improvement and Effective Planning, Goal / QPC, Methuen, MA, 1994.
146. Fitch K, Bernstein SJ, Aguilar MD, Burnand B, LaCalle JR, Lázaro P, van het Loo M, McDonnell J, Vader JP, Kahan JP. The RAND/UCLA Appropriateness Method User's Manual. RAND Corporation, Santa Monica, CA, 2001.
147. Commission on Health Research for Development. Health Research: Essential Link to Equity in Development. Oxford University Press, NY, 1990.
148. Yoshida S. Approaches, tools and methods used for setting priorities in health research in the 21(st) century. J Glob Health. 2016;6(1):010507.

149. GRADE Working Group. Grading quality of evidence and strength of recommendations. Br Med J. 2004;328:1490–1494.
150. Moberg J, Oxman AD, Rosenbaum S, Schünemann HJ, Guyatt G, Flottorp S, Glenton C, Lewin S, Morelli A, Rada G, Alonso-Coello P; GRADE Working Group. The GRADE Evidence to Decision (EtD) framework for health system and public health decisions. Health Res Policy Syst. 2018 May 29;16(1):45.
151. Chalmers I, Bracken MB, Djulbegovic B, Garattini S, Grant J, Gülmezoglu AM, Howells DW, Ioannidis JPA, Oliver, S. How to increase value and reduce waste when research priorities are set. Lancet 2014; 383:156–165.
152. Global Forum for Health Research. The 10/90 Report on Health Research 2003-2004. Global Forum for Health Research, Geneva, 2004.
153. Ghaffar A, de Francisco A, Matlin S. The combined approach matrix: a priority-setting tool for health research. Global Forum for Health Research, Geneva, 2004.
154. Ghaffar A. Setting research priorities by applying the combined approach matrix. Indian J Med Res. 2009;129:368-75.
155. Song P, Adeloye D, Acharya Y, Bojude DA, Ali S, Alibudbud R, Bastien S, Becerra-Posada F, Berecki M, Bodomo A, Borrescio-Higa F, Buchtova M, Campbell H, Chan KY, Cheema S, Chopra M, Cipta DA, Castro LD, Ganasegeran K, Gebre T, Glasnović A, Graham CJ, Igwesi-Chidobe C, Iversen PO, Jadoon B, Lanza G, Macdonald C, Park C, Islam MM, Mshelia S, Nair H, Ng ZX, Htay MNN, Akinyemi KO, Parisi M, Patel S, Peprah P, Polasek O, Riha R, Rotarou ES, Sacks E, Sharov K, Stankov S, Supriyatiningsih W, Sutan R, Tomlinson M, Tsai AC, Tsimpida D, Vento S, Glasnović JV, Vokey LBV, Wang L, Wazny K, Xu J, Yoshida S, Zhang Y, Cao J, Zhu Y, Sheikh A, Rudan I; International Society of Global Health (ISoGH). Setting research priorities for global pandemic preparedness: An international consensus and comparison with ChatGPT's output. J Glob Health. 2024;14:04054.
156. Myers SC. Determinants of corporate borrowing. J Financ Econ, 1977; 5:147-175.
157. Dixit AK, Pindyck RS. Investment under Uncertainty. Princeton University Press, Princeton, NJ, 1994.
158. Trigeorgis L. Real Options Managerial Flexibility and Strategy in Resource Allocation. MIT Press, Cambridge, MA, 1996.
159. Henderson BD. The Product Portfolio. Boston Consulting Group's Perspectives, No. 66. Boston, MA, 1970.
160. McKinsey and Company. Enduring Ideas: The GE–McKinsey nine-box matrix. McKinsey Quarterly, New York, NY, 2008.
161. Wind Y, Mahajan V, Swire DJ. An empirical comparison of standardized portfolio models. J Marketing 1983; 47:89–99.
162. Phaal R, Farrukh CJP, Probert DR. Technology roadmapping - a planning framework for evolution and revolution. Technol Forecast Soc Change 2004; 71:5–26.
163. Kano N, Seraku N, Takahashi F, Tsuji S. Attractive quality and must-be quality. J Jap Soc Qual Control 1984; 14:39-48.
164. Cooper RG. Stage-gate systems: a new tool for managing new products. Business Horizons, 1990;33:44-54.
165. Collins AM, Quillian MR. Retrieval time from semantic memory. J Verb Learn Verb Behav 1969; 8:240-247.
166. Knowledge Graph (Google). Available from: https://en.wikipedia.org/wiki/Knowledge_Graph_(Google); Accessed: 27 Jul 2025;
167. Unified Medical Language System (UMLS). Available from: https://en.wikipedia.org/wiki/Unified_Medical_Language_System; Accessed: 27 Jul 2025;
168. Percha B, Altman RB. Learning the structure of biomedical relationships from unstructured text. PLOS Computational Biology 2015; 11:e1004513.
169. Himmelstein DS, Lizee A, Hessler C, Brueggeman L, Chen SL, Hadley D, Green A, Khankhanian P, Baranzini SE. Systematic integration of biomedical knowledge prioritizes drugs for repurposing. Elife. 2017; 6:e26726.
170. IBM Watson. Available from: https://en.wikipedia.org/wiki/IBM_Watson; Accessed: 27 Jul 2025;
171. Semantic Scholar. Availablef rom: https://en.wikipedia.org/wiki/Semantic_Scholar; Accessed: 27 Jul 2025;
172. https://en.wikipedia.org/wiki/Comparison_of_research_networking_tools_and_research_profiling_systems; Accessed: 27 Jul 2025;
173. Microsoft Academic Graph. Available from: https://www.microsoft.com/en-us/research/project/microsoft-academic-graph/; Accessed: 27 Jul 2025;
174. National Center for Advancing Translational Sciences. Biomedical Data Translator Consortium. Available from: https://ncats.nih.gov/research/research-activities/translator; Accessed: 27 Jul 2025;
175. Ehrlinger L, Wöß W. Towards a definition of knowledge graphs, 2016. In: SEMANTICS, CEUR Workshop Proceedings, CEUR-WS.org, Vol 1695.

176. Wang Q, Mao Z, Wang B, Guo L. Knowledge graph embedding: A survey of approaches and applications. IEEE Transactions on Knowledge and Data Engineering 2017;29:2724–2743.
177. Percha B, Altman RB. Learning the Structure of Biomedical Relationships from Unstructured Text. PLoS Comput Biol. 2015;11:e1004216.
178. https://en.wikipedia.org/wiki/Artificial_intelligence; Accessed: 27 Jul 2025;
179. Sutton RS, Barto AG. Reinforcement Learning: An Introduction. MIT Press, Cambridge, MA, 1998.
180. Moody J, Saffell M. Learning to trade via direct reinforcement, IEEE Trans. Neural Networks, 2001; 12:875–889.
181. Jiang Z, Xu D, Liang J. A deep reinforcement learning framework for the financial portfolio management problem. arXiv 2017; Available from: https://arxiv.org/abs/1706.10059; Accessed: 27 Jul 2025.
182. Alpha Zero. Available from: https://en.wikipedia.org/wiki/AlphaZero; Accessed: 27 Jul 2025.
183. Garry J, Tomlinson M, Lohan M. The potential role of AI in research priority setting exercises. J Glob Health. 2025 Jun 6;15:03019.
184. Chan Zuckerberg Initiative. Available from: https://chanzuckerberg.com; Accessed: 25 Jul 2025;
185. Park YJ, Kaplan D, Ren Z, Hsu CW, Li C, Xu H, Li S, Li J. Can ChatGPT be used to generate scientific hypotheses? J Materiomics 2024; 10:578-584.
186. Fricke S. Semantic Scholar. J Med Libr Assoc. 2018;106:145–7.
187. Lakatos I. The Methodology of Scientific Research Programmes: Philosophical Papers, Volume 1. Cambridge University Press, Cambridge, 1978.
188. Feyerabend P. Against Method: Outline of an Anarchistic Theory of Knowledge. New Left Books, London, 1975.
189. Engel GL. The need for a new medical model: a challenge for biomedicine. Science 1977; 196:129-36.
190. Longino H. Science as Social Knowledge: Values and Objectivity in Scientific Inquiry. Princeton University Press, Princeton, NJ, 1990.
191. Funtowicz S, Ravetz J. Science for the Post-Normal Age. Futures, 25:739–755, 1993.
192. Nowotny H, Scott P, Gibbons M. Re-thinking science: Knowledge and the public in an age of uncertainty. Polity Press, Cambridge, 2001.
193. Cartwright N. Hunting Causes and Using Them: Approaches in Philosophy and Economics. Cambridge University Press, Cambridge, 2007.
194. Crosby N. Citizen juries: One solution for difficult environmental questions. In: Renn O, Webler T, Wiedemann P (Eds). Fairness and Competence in Citizen Participation: Evaluating Models for Environmental Discourse. Kluwer Academic Publishers, Dordrecht, 1995., pp.157-171.
195. Dryzek JS. Deliberative Democracy and Beyond: Liberals, Critics, Contestations. Oxford University Press, Oxford, 2000.
196. Fishkin JS. Democracy and Deliberation: New Directions for Democratic Reform. Yale University Press, New Haven, CT, 1991.
197. Fishkin JS, Luskin RC. Experimenting with a democratic ideal: Deliberative polling and public opinion. Acta Politica 2005; 40:284–298.
198. Local Government association. Rules: The Gunning Principles. Available from: https://www.local.gov.uk/sites/default/files/documents/The%20Gunning%20Principles.pdf; Accessed: 27 Jul 2025;
199. Kiesler S, Sproull L. Response effects in the electronic survey. Publ Opin Quarterly 1986; 50:402-413.
200. OECD. Citizens as Partners: OECD Handbook on Information, Consultation and Public Participation in Policy-Making. OECD Publishing, Paris, 2001.
201. National Institute for Health and Care Excellence. Guide to the Methods of Technology Appraisal. NICE, London, 2004.
202. Cuhls K, Popper R. Foresight and public policy in Europe. Int J Foresight Innov Policy 2008; 4:106–120.
203. Santos BS. Participatory budgeting in Porto Alegre: Toward a redistributive democracy. Politics and Society 1998; 26(4):461-510.
204. Baiocchi G. Radicals in power: The Workers' Party and experiments in urban democracy in Brazil. Zed Books, London, 2003.
205. Wampler B. Participatory budgeting in Brazil: Contestation, cooperation, and accountability. Pennsylvania State University Press, University Park, PA, 2007.
206. Decidim platform. Available from: https://en.wikipedia.org/wiki/Decidim; Accessed: 27 Jul 2025;
207. Consul platform. Available from: https://en.wikipedia.org/wiki/Consul_(software); Accessed: 27 Jul 2025;
208. Ockham, W. Ockham's Theory of Terms, Part I of the Summa Logicae, circa 1323. Translated and introduced by Loux MJ, University of Notre Dame Press, Notre Dame, IN, 1974.

209. Ockham, W. Quodlibetal Questions. circa 1328. Translated by Freddoso AJ, Kelley FE, Yale University Press, New Haven, CT, 1978.
210. Hamilton W. Discussions on Philosophy, Literature and Education. William Blackwood and Sons, Edinburgh, 1852.
211. Akaike H. A new look at the statistical model identification. IEEE Transactions on Automatic Control, 1974; 19:716-723.
212. Schwarz G. Estimating the dimension of a model. The Annals of Statistics 1978; 6:461-464.
213. Bayes T. An Essay towards Solving a Problem in the Doctrine of Chances. Philosophical Transactions of the Royal Society of London 1763; 53:370–418.
214. Laplace PS. Essai philosophique sur les probabilités. Mme. Ve. Courcier, Paris, 1814.
215. Hegel GWF. Phänomenologie des Geistes. Frommann-Holzboog, Bamberg, 1807.
216. Chalybäus HM. Historische Entwicklung der spekulativen Philosophie von Kant bis Hegel. Arnoldi, Dresden, 1837.
217. West TG. Plato's Apology of Socrates: An Interpretation, with a New Translation. Cornell University Press, Ithaca, NY, 1979.
218. Mill JS. On Liberty. Publisher. Longman, Roberts, and Green Co, London, 1859.
219. Simpson CR, Robertson C, Vasileiou E, McMenamin J, Gunson R, Ritchie LD, Woolhouse M, Morrice L, Kelly D, Stagg HR, Marques D, Murray J, Sheikh A. Early Pandemic Evaluation and Enhanced Surveillance of COVID-19 (EAVE II): protocol for an observational study using linked Scottish national data. BMJ Open. 2020;10(6):e039097.
220. Vasileiou E, Simpson CR, Shi T, Kerr S, Agrawal U, Akbari A, Bedston S, Beggs J, Bradley D, Chuter A, de Lusignan S, Docherty AB, Ford D, Hobbs FR, Joy M, Katikireddi SV, Marple J, McCowan C, McGagh D, McMenamin J, Moore E, Murray JL, Pan J, Ritchie L, Shah SA, Stock S, Torabi F, Tsang RS, Wood R, Woolhouse M, Robertson C, Sheikh A. Interim findings from first-dose mass COVID-19 vaccination roll-out and COVID-19 hospital admissions in Scotland: a national prospective cohort study. Lancet. 2021;397(10285):1646-1657.
221. Negrini S, Ceravolo MG, Côté P, Arienti C. A systematic review that is "rapid'' and "living'': A specific answer to the COVID-19 pandemic. J Clin Epidemiol. 2021;138:194-198.
222. Ashcroft T, McSwiggan E, Agyei-Manu E, Nundy M, Atkins N, Kirkwood JR, Ben Salem Machiri M, Vardhan V, Lee B, Kubat E, Ravishankar S, Krishan P, De Silva U, Iyahen EO, Rostron J, Zawiejska A, Ogarrio K, Harikar M, Chishty S, Mureyi D, Evans B, Duval D, Carville S, Brini S, Hill J, Qureshi M, Simmons Z, Lyell I, Kavoi T, Dozier M, Curry G, Ordóñez-Mena JM, de Lusignan S, Sheikh A, Theodoratou E, McQuillan R. Effectiveness of non-pharmaceutical interventions as implemented in the UK during the COVID-19 pandemic: a rapid review. J Public Health (Oxf). 2025;47:268-302.
223. Systematic Review. Available from: https://en.wikipedia.org/wiki/Systematic_review; Accessed: 25 Jul 2025.
224. Millum J. Ethics and health research priority setting: a narrative review. Wellcome Open Res. 2024;9:203.
225. Empirical Research. Available from: https://en.wikipedia.org/wiki/Empirical_research; Accessed: 25 Jul 2025.
226. Johnson S. Where Good Ideas Come From. Penguin Books, London, 2010.
227. Page MJ, McKenzie JE, Bossuyt PM, Boutron I, Hoffmann TC, Mulrow CD, Shamseer L, Tetzlaff JM, Akl EA, Brennan SE, Chou R, Glanville J, Grimshaw JM, Hróbjartsson A, Lalu MM, Li T, Loder EW, Mayo-Wilson E, McDonald S, McGuinness LA, Stewart LA, Thomas J, Tricco AC, Welch VA, Whiting P, Moher D. The PRISMA 2020 statement: an updated guideline for reporting systematic reviews. BMJ. 2021;372:n71.
228. Hopewell S, Chan AW, Collins GS, Hróbjartsson A, Moher D, Schulz KF, Tunn R, Aggarwal R, Berkwits M, Berlin JA, Bhandari N, Butcher NJ, Campbell MK, Chidebe RCW, Elbourne D, Farmer A, Fergusson DA, Golub RM, Goodman SN, Hoffmann TC, Ioannidis JPA, Kahan BC, Knowles RL, Lamb SE, Lewis S, Loder E, Offringa M, Ravaud P, Richards DP, Rockhold FW, Schriger DL, Siegfried NL, Staniszewska S, Taylor RS, Thabane L, Torgerson D, Vohra S, White IR, Boutron I. CONSORT 2025 statement: updated guideline for reporting randomised trials. BMJ. 2025;389:e081123.